\documentclass[apj]{emulateapj}
\usepackage{natbib}
\usepackage[breaklinks=true]{hyperref}
\usepackage{rotating}

\newcommand\footnoterule{\kern -2pt \hrule width 2in \kern 2.6pt}
\usepackage{color}
\definecolor{myred}{rgb}{0.8,0.0,0.0}
\definecolor{myblue}{rgb}{0.2,0.0,0.7}
\definecolor{mybrown}{rgb}{0.5,0.2,0.0}

\newcommand\hide[1]{}
\slugcomment{Submitted to ApJ January 13, 2015; Accepted March 5, 2015}

\shorttitle{A young planetary mass companion VHS\,1256-1257~b}
\shortauthors{B. Gauza et al.}

\begin{document}


\title{Discovery of a young planetary mass companion to the nearby M~dwarf VHS~J125601.92-125723.9\,$^*$}


\author{
Bartosz Gauza\altaffilmark{1,2},  
Victor~J.~S. B\'ejar\altaffilmark{1,2}, 
Antonio P\'erez-Garrido\altaffilmark{3},  
Maria Rosa Zapatero Osorio\altaffilmark{4},\\ 
Nicolas Lodieu\altaffilmark{1,2}, 
Rafael Rebolo\altaffilmark{1,2,5}, 
Enric Pall\'e\altaffilmark{1,2} and
Grzegorz Nowak\altaffilmark{1,2}
}
\altaffiltext{*}{Based on observations collected at the European Southern Observatory, Chile, programs 092.C-0874 and 293.C-5014(A).}
\altaffiltext{1}{Instituto de Astrof\'isica de Canarias (IAC), Calle V\'ia L\'actea s/n, E-38200 La Laguna, Tenerife, Spain}
\altaffiltext{2}{Departamento de Astrof\'isica, Universidad de La Laguna (ULL), E-38205 La Laguna, Tenerife, Spain}
\altaffiltext{3}{Dpto. F\'isica Aplicada, Universidad Polit\'ecnica de Cartagena, Campus Muralla del Mar, Cartagena, Murcia E-30202, Spain}
\altaffiltext{4}{Centro de Astrobiolog\'ia (CSIC-INTA), Ctra. Ajalvir km 4, 28850, Torrej\'on de Ardoz, Madrid, Spain}
\altaffiltext{5}{Consejo Superior de Investigaciones Cient\'ificas, CSIC, Spain}

\email{bgauza@iac.es}

\begin{abstract}
\hspace*{-3.2mm}In a search for common proper motion companions using the VISTA Hemisphere Survey (VHS) and the 2MASS catalogs we 
have identified a very red ($J-K_s=2.47$~mag) late-L dwarf companion 
of a previously unrecognized M dwarf VHS\,J125601.92-125723.9 (hereafter VHS\,1256-1257), located  
at a projected angular separation of 8\farcs06\,$\pm$\,0\farcs03. In this work we 
present a suite of astrometric, photometric, and spectroscopic observations of this new 
pair in an effort to confirm the companionship and characterize the components. From 
low-resolution ($R$\,$\sim$\,130--600) optical and near-infrared spectroscopy we classified 
the primary and the companion as an M$7.5\pm0.5$ and L$7\pm1.5$, respectively. 
The primary shows slightly weaker alkali lines than field dwarfs of similar spectral type,
but still consistent with either a high-gravity dwarf or a younger object of hundreds of millions of years.
The secondary shows spectral features characteristic for low surface gravity objects at ages below several hundred million years,
like the peaked triangular shape of the $H$-band continuum and alkali lines weaker than in field dwarfs of the same spectral type.
The absence of lithium in the atmosphere of the primary and the likely kinematic membership to the Local Association 
allowed us to constrain the age of the system to the 
range of 150--300 Myr. We report a measurement of the trigonometric 
parallax $\pi$\,=\,78.8\,$\pm$\,6.4~mas, which translates into a distance of 12.7\,$\pm$\,1.0~pc; 
the pair thus has a projected physical separation of 102\,$\pm$\,9~AU. We derived the bolometric 
luminosities of the components and compared them with theoretical evolutionary 
models to estimate the masses and effective temperatures. For the primary, we determined a
luminosity of $\log(L_{\rm bol}/L_{\odot}) = -3.14\pm0.10$, and inferred a mass of 73$^{+20}_{-15}$ $M_{\rm Jup}$ 
at the boundary between stars and brown dwarfs and an effective temperature of $2620\pm140$ K.
For the companion we obtained a luminosity of $\log(L_{\rm bol}/L_{\odot}) = -5.05\pm0.22$ and a mass of 11.2$^{+9.7}_{-1.8}$ $M_{\rm Jup}$ 
placing it near the deuterium-burning mass limit. The effective temperature derived from evolutionary models is 880$^{+140}_{-110}$ K, 
about 400--700 K cooler than the temperature expected for field late-L dwarfs.
\end{abstract}


\section{Introduction}
Very low-mass stars and brown dwarfs directly imaged around stars constitute an
important group for studies of the properties of substellar objects. We can infer the distance 
and metallicity of the companion from the brighter, more easily characterizable primary star. 
Most importantly, we can constrain the age of the system and thus overcome the intrinsic 
degeneracy between mass and age for the temperature and luminosity of objects below the hydrogen 
burning limit. This allows for a more thorough characterization and offers the opportunity to
better understand their physical properties and to test evolutionary and atmospheric models 
(e.g., \citealt{2006MNRAS.368.1281P},\linebreak \citealt{2010AJ....139..176F}).
There are about 1500 objects spectrally 
classified as M7 and later, the vast majority is found to be isolated, single sources that appear 
to have spectral energy distributions, photometric colors, and kinematics consistent with a field 
population with ages in the range from 2 to 8 Gyr \citep{2004AJ....127.3553K, 2007AJ....133..439C, 
2009AJ....137....1F}. About 120 ultracool dwarfs (late-M, L, T, and Y dwarfs) have been  
confirmed as components of binary or multiple systems \citep{2010AJ....139..176F, 
2014ApJ...792..119D}. 
%
%
These objects discovered over the last two decades have exhibited a large diversity in ages, 
atmospheric properties, and chemical compositions.

\begin{figure*}
  \centering
  \includegraphics[scale=0.76,keepaspectratio=true]{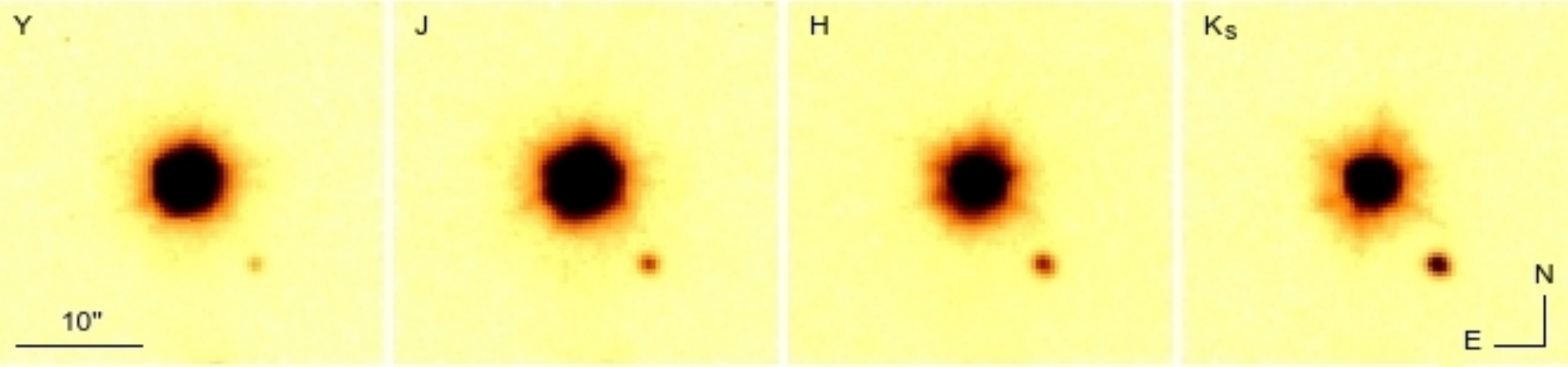}
  \caption{VISTA $YJHK_s$ band images of the new common proper motion pair VHS\,1256-1257. The 
  identified companion is located at a position angle of 218\fdg1$\pm$0\fdg2, with a projected 
  angular separation of 8\farcs06$\pm$0\farcs03 which corresponds to $\sim102$ AU at the determined 
  parallactic distance of 12.7~pc. The field of view is 30\arcsec$\times$30\arcsec, with north up and east 
  to the left.}
  \label{vhsimages}
\end{figure*}

Surveys aimed at identifying low-mass companions of young stars \citep[e.g.,][]{2011ApJ...729..139W, 
2013ApJ...774...55B, 2014arXiv1405.1560C}, and the least massive members of young star clusters 
\citep[e.g.,][]{2011ApJ...743...64B, 2012ApJ...754...30P, 2014A&A...568A..77Z}, as well as 
searches for L and T dwarfs in the field using large sky area surveys \citep{2006ApJ...639.1120K, 
2009AJ....137.3345C} have revealed a number of ultracool dwarfs that exhibit photometric 
and spectral properties different from typical late-type objects of the field population. Some of 
these peculiarities have been attributed to low surface gravities and cloudy atmospheres, occurring 
at the early stages of evolution, below several hundred million years 
\citep{2009AJ....137.3345C, 2013ApJ...772...79A, 2014A&A...568A...6Z}. 
Young L dwarfs, either free-floating or as companions to stars, were 
found to share similar characteristics.
They have very red colors ($J-K_s$\,$>$\,2~mag), $J$-band absolute magnitudes fainter than their old, 
field counterparts, and show distinctive spectral features such as, for example, a sharply peaked, 
triangal-shaped continuum in the $H$ band and weaker sodium and potassium lines.

Recent studies have revealed a strong resemblance between the young L dwarfs and directly 
imaged planetary mass companions (e.g., 2MASS 1207-39\,b, \citealt{2005A&A...438L..25C}, HR 8799\,bcde, 
\citealt{2008Sci...322.1348M, 2010Natur.468.1080M}, GJ 504\,b \citealt{2013ApJ...774...11K}, 
$\beta$ Pic b, \citealt{2009A&A...493L..21L}).
They have similar near-infrared (near-IR) colors and absolute magnitudes, overlapping effective 
temperature regimes of $\sim$1000--1500 K and masses of a few to a few tens of Jupiter masses 
\citep{2011ApJ...733...65B, 2013ApJ...774...55B, 2013AJ....145....2F, 2014IAUS..299...36F, 2013ApJ...777L..20L}. 
Spectroscopic and photometric studies of young substellar objects can provide information 
on the physical properties of gas giant exoplanets found by transit and radial velocity surveys, 
in particular, on the characteristics and composition of their complex atmospheres.
 
In this work we present the identification and characterization of a nearby, young binary
system with components that we classify as M$7.5\pm0.5$ and L$7\pm1.5$. In Section 2
we describe the data and search method that led to the identification of the pair.
Section 3 contains the description of observations aimed at confirming their companionship 
and characterization. In Section 4 we discuss their physical properties. 
We classify their spectral types, and determine the distance and radial and space velocities.
We constrain the possible age of the system and estimate the luminosities, masses, and
effective temperatures of the two components. Final remarks and conclusions are presented in Section 5.

\section{Identification of the system} 
\subsection{VISTA Hemisphere Survey Data}
The reported common proper motion pair was identified using the VISTA Hemisphere Survey (VHS) data
and the 2MASS \citep{2006AJ....131.1163S}.
The VHS is a near-IR ESO public survey designed to map the entire southern hemisphere
in the $J$ and $K_s$ broad-band filters with average 5$\sigma$ depths of $J = 20.2$~mag 
and $K_s = 18.1$~mag \citep{2013Msngr.154...35M}. In some particular areas also observations in the $Y$ and $H$ bands 
are also performed. The 4-m VISTA telescope \citep{2001ASPC..232..339E, 
2004Msngr.117...27E} has operated since 2009 at ESO's Cerro Paranal Observatory in Chile and has thus
far covered about 8000 deg$^2$ of the sky. It is equipped with a wide-field infrared 
camera VIRCAM \citep{2006SPIE.6269E..30D} composed of 16 Raytheon detectors 2048$\times$2048
pixel array each, with a mean plate scale of 0\farcs34, giving a field of view of 1.65$\degr$ in diameter.

The VHS images are processed and calibrated automatically by a dedicated science pipeline 
implemented by the Cambridge Astronomical Survey Unit (CASU). Standard reduction and 
processing steps include dark and sky subtraction, flat-field correction, linearity 
correction, destripe, and jitter stacking. For a detailed description we refer the reader 
to the CASU webpage \url{http://casu.ast.cam.ac.uk/surveys-projects/vista} as well as to 
\cite{2004SPIE.5493..411I} and \cite{2010ASPC..434...91L}. 

The photometry provided in the VHS catalog is calibrated using the 2MASS magnitudes of 
all matching stars converted onto the VISTA system applying color equations\footnote[6]
{\url{http://casu.ast.cam.ac.uk/surveys-projects/vista/technical/photometric-properties}}, 
which include terms accounting for the interstellar reddening. Photometric calibrations are 
determined to an accuracy of 1\%--2\%. The astrometric solution for VHS observations is obtained
through the CASU pipeline, using the 2MASS point source catalog. The objects on the catalogs
extracted from each detector are matched to their counterparts in 2MASS using a correlation
radius of 1\arcsec. Because 2MASS has a high degree of internal consistency it is possible
to calibrate the world coordinate system of VISTA images to relative accuracy better than 0\farcs1.

\subsection{Search Method}
Using VHS data, we carried out a search for high proper motion objects, by cross-matching VHS 
sources with the 2MASS Point Source Catalog \citep{2006AJ....131.1163S}. The search focused on objects that had moved 
at least 2\arcsec and a maximum of 30\arcsec from 2MASS to VHS. The time baseline between the 
two surveys is typically about 12~yr, which gives proper motions of approximately 0.15--3.0 arcsec\,yr$^{-1}$.
Over the common area of $\sim$5000 deg$^2$ between 2MASS and VHS we found more than 6000 objects
with $J$ magnitudes in the range 11--17\,mag and proper motion higher than 150~mas\,yr$^{-1}$. We have cross-correlated
this catalog with {\it WISE} \citep{2010AJ....140.1868W}, so that near and mid-infrared information is
available for each target. Most of the high proper motion objects are relatively nearby M dwarfs with estimated
photometric distances $<$50~pc (A. Perez-Garrido et al., in preparation). We have searched for common proper motion pairs and multiples among these objects with
proper motion consistent within 40~mas\,yr$^{-1}$ in both right ascension and declination ($\mu_\alpha$, $\mu_\delta$). 
VHS\,1256-1257 system was one of the identified candidates. 
The pair was observed with the VISTA $YJHK_s$ filters and with a seeing of 1\farcs1 on 2011 July 1. 
The primary and the secondary were cataloged in 2MASS with designations 2MASS\,J125602.15-125721.7 and 2MASS\,J125601.83-125727.6, respectively.
The primary was also listed as SIPS\,1256-1257 in the sample of low-mass stars with $\mu>0\farcs1\,$yr$^{-1}$ from \cite{2007A&A...468..163D},
with measured proper motion amplitude $\mu$\,=\,0\farcs357\,yr$^{-1}$ and the position angle of the proper motion vector of 244.76\degr.

\begin{figure}
\centering
 \includegraphics[scale=0.72,keepaspectratio=true]{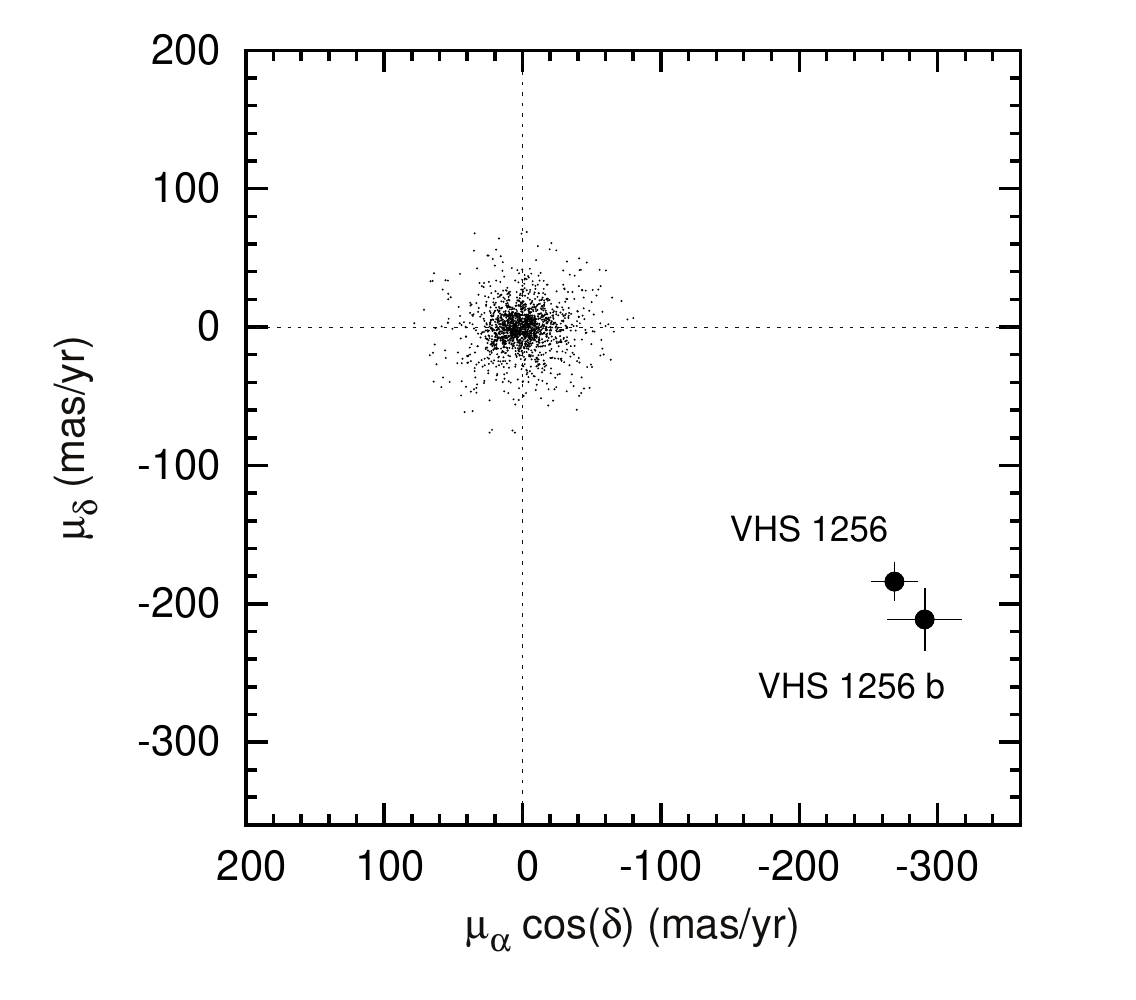}
  \caption{Proper motion diagram for the VHS\,1256-1257 system. All correlated 
  objects within one degree around the primary with $J<$17.5\,mag are plotted with black dots.
  Components of the common proper motion pair are labeled. Error bars correspond to the 
  astrometric RMS of stars near the magnitudes of the components.}
\label{pmdiag}
\end{figure}

VISTA $YJHK_s$ images of the pair are presented in Fig.\,\ref{vhsimages} and the photometry is given in Table \ref{measurements}.
The companion is located at a projected angular separation of 8\farcs06$\pm$0\farcs03, at
a position angle of 218\fdg1$\pm$0\fdg2. This separation corresponds to a projected orbital separation
of $102\pm9$ AU at the estimated distance of the system (Section 4.2).
The two components share a common proper motion, which significantly differs from the proper motion of background stars as 
shown in Fig.\,\ref{pmdiag}. 
The $\mu_{\alpha}\cos\delta$ and $\mu_{\delta}$ measured from the VHS and 2MASS positions of the sources, 
were $-270\pm17$,~$-185\pm14$ and $-292\pm27$ and $-212\pm23$ mas\,yr$^{-1}$,
for the primary and secondary, respectively. 
The uncertainties in proper motion of each component correspond to the rms of $\mu_{\alpha}\cos\delta$ and $\mu_{\delta}$ of 
the background stars within a one-degree radius around the primary, and magnitudes similar to that of
the given component. Time baseline between the two epochs is 12.3~yr.
The primary has $J=11.02\pm0.02$ mag and $J-K_s=0.97$ mag 
(2MASS photometry, VIRCAM is out of the linear range) 
The secondary is roughly six magnitudes fainter in the $J$ band ($J=17.14\pm0.02$ mag) and
has a very red $J-K_s$ color of 2.47 mag indicating its significantly cooler type and lower 
mass with respect to the primary.

\begin{table*}
\centering
\scriptsize
\caption{Observation log of VHS\,1256-1257.\label{log}}
\begin{tabular*}{\textwidth}{l @{\extracolsep{\fill}} l c c c c c c c c}
\hline
\hline
Obs. Date & Tel/Instrument & Mode    & Wavelength  & Exp. Time  & Seeing    &  Slit     & Grating & Scale          & Res. \\     
(UT)      &                &         & ($\mu$m) & (s)        & (\arcsec) & (\arcsec) &         & (\arcsec/pix)  & Power\\                            
\hline
2011 Jul 01\tablenotemark{a}         & VISTA/VIRCAM & Img       & $YJHK_s$    & 15, 15, 7.5, 7.5     & 1.0 & -- & --  & 0.34 & -- \\                          
2014 Mar 02\tablenotemark{b}        & GTC/OSIRIS   & Spec   & 0.48--1.00  & 120 &                0.57      & 1.5 & R500R & 0.25 & 320 \\                                  
2014 Mar 12\tablenotemark{a}        & NTT/SofI     & Spec      & 0.95--1.64  & 4$\times$600  &  0.8 & 1.0 & Blue & 0.29 & 600\\                
2014 Mar 12\tablenotemark{a}        & NTT/SofI     & Spec      & 1.53--2.52  & 4$\times$900 &  0.8 & 1.0 & Red & 0.29 & 600 \\                
2014 Apr 22, May 04\tablenotemark{b} & NOT/ALFOSC   & Spec   & 0.63--0.68  & 900, 4$\times$900  & 0.8--1.1 & 1.3 & Grism\#17 & 0.19 & 4700\\                              
2014 Apr 27\tablenotemark{a}       & NTT/SofI     & Img       & $JH$        & 9$\times$2 & 1.0--1.3 & -- & -- & 0.29 & -- \\
2014 May 14\tablenotemark{b}          & GTC/OSIRIS   & Spec   & 0.56--0.77  & 3$\times$600  &  0.56 &  0.6 & R2500R & 0.25 & 3000\\                  
2014 Jun 03\tablenotemark{a}        & GTC/OSIRIS   & Spec    & 0.48--1.00  & 3$\times$600  & 0.63 & 1.5 & R300R & 0.25 & 130  \\                
2014 Jul 01, 09, 10\tablenotemark{b} & VLT/UVES     & Spec   & 0.37--0.49  & 3$\times$3300 &  0.5--1.7 &  1.0  & CD2, 4 & 0.22 & 40000\\                         
2014 Jul 01, 09, 10\tablenotemark{b} & VLT/UVES     & Spec   & 0.56--0.95  & 3$\times$3300 &  0.5--1.7 &  1.0  & CD2, 4 & 0.16 & 40000\\                         
2014 Jul 15\tablenotemark{b}         & IAC80/CAMELOT& Img    & $VI$                       & 300 & 1.1 & -- & -- & 0.30 & -- \\                           
2014 Jul 17\tablenotemark{a}        & WHT/ACAM     & Img      & Sloan $i, z$ & 9$\times$30, 9$\times$30 & 0.85  & -- & -- & 0.25 & --\\   
2014 Jul 18\tablenotemark{a}         & WHT/LIRIS    & Img     & $J$          & 5$\times$9$\times$2 & 0.7  & -- & -- & 0.25 & -- \\   
2014 Dec 19, 20\tablenotemark{b} & IAC80/CAMELOT& Img   & $I$                       & 6$\times$300 & 1.5, 1.2 & -- & -- & 0.30 & --\\                           
\hline
\end{tabular*}
\tablenotetext{1}{ Both components were observed} 
\tablenotetext{2}{ Only the primary was observed}
\end{table*}

\section{Follow-up observations and data reduction}
\subsection{NTT/SofI Near-infrared Spectroscopy and Imaging}
To measure the infrared spectral types of the VHS\,1256-1257 components we performed 
follow-up near-IR spectroscopy using the Son of ISAAC (SofI) spectro-imager 
installed on the 3.6\,m New Technology Telescope (NTT) on 2014 March 12. SofI is 
equipped with a Hawaii HgCdTe detector with 1024$\times$1024 pixel array offering a 
field of view of 4.9$\times$4.9 arcmin$^2$ with a 0\farcs288 pixel scale. 
We have used blue and red grisms covering the 950--1640 and 
1530--2520\,nm range, combined with a slit of 1\arcsec, orientated along the position 
angle of the components of the system. This instrumental configuration provides a 
nominal dispersion of 6.96 and 10.22~\AA\,pix$^{-1}$ for the Blue and Red grating, respectively. 
To subtract the sky background we used an ABBA nodding pattern with 
an offset of 20\arcsec between the two positions. Individual exposure times were 
600 and 900\,s for the red and blue grism, respectively. Right after VHS\,1256-1257 we 
observed a telluric standard star HD\,112304 \citep[$V=6.19$\,mag, A0V,][]{2000AA...355L..27H, 2007A&A...474..653V}. 
The sky conditions during the observations were clear with sub-arcsecond seeing.

The two-dimensional (2D) images were reduced and combined using the ESO SofI pipeline run within the $Gasgano$ environment.
The traces of the primary and the companion spectra on the 2D images were clearly separated and there was no contamination between the two sources. 
The two spectra of the VHS~1256-1257 system and of HD\,112304 were then extracted using standard IRAF routines within the {\sc apall} package
and wavelength calibrated via Xenon arc lines. The dispersion solution had an rms of 0.4~and 0.35\,\AA~for the blue and red part of the spectrum, respectively. The final resolution of the spectrum was 
24\,\AA~($R$\,$\sim$\,600). Telluric absorption lines were corrected, dividing the spectra by the A0V telluric standard HD\,112304 
observed at a similar airmass, and multiplying by a blackbody of a corresponding 
effective temperature of 9480\,K. The obtained NTT SofI spectra of VHS\,1256-1257 primary and companion 
normalized at 1.6\,$\mu$m are displayed in Figure\,\ref{fullspec}. 

On 2014 April 27 we used SofI in imaging mode, to perform $J$- and $H$-band
observations at an additional epoch. We used a nine-position dither pattern with individual
exposures of 5\,s. The weather was clear, with a seeing of 1\farcs15--1\farcs30.
Raw images were reduced using the ESO SofI Pipeline within {\it Gasgano}, which includes bias subtraction,
flat-field correction, plane re-centering, and co-addition of individual frames to a final image.
Images were employed for determining the trigonometric parallax of the system and
for performing relative astrometry between the two components as described in Section 4.2.
Table \ref{par_observations} provides these measurements.

\subsection{GTC/OSIRIS Optical Spectroscopy}
We obtained low-resolution optical spectroscopy of VHS\,1256-1257 with the Optical System for Imaging and 
low-intermediate Resolution Integrated Spectroscopy 
\citep[OSIRIS;][]{2010hsa5.conf...15C}
mounted on the 10.4-m Gran Telescopio de Canarias (GTC) telescope in La Palma. OSIRIS is equipped with two 2048$\times$4096 
Marconi CCD42-82 detectors offering a field of view approximately 7$\times$7 arcmin$^2$ with an 
unbinned pixel scale of 0\farcs125. 
We observed the primary star using two different gratings: R500R, which allowed us to measure the general spectral 
energy distribution at a resolution of $R$\,$\sim$\,320, covering the 0.48--1.00 $\mu$m range, and R2500R providing a resolution 
of $R$\,$\sim$\,3000 in the 0.56--0.77 $\mu$m region, which we used to study the H$\alpha$ and Li\,{\sc i} lines. Observations 
with R500R and R2500R were acquired on 2014 March 2 and May 14, respectively. On 2014 June 3, both the primary and
secondary were aligned on the 1\farcs5-width slit and were observed simultaneously using the R300R grating. This
instrumental setup provided optical spectra with a resolving power of $R$\,$\sim$\,130, which is optimal to maximize
the signal-to-noise ratio of the faint companion data.
Observations were performed in service mode (Table \ref{log}) as part of a 
GTC filler program (number GTC5-14A; PI Lodieu). The nights 
were dark with a clear sky and a seeing of $\sim$0\farcs6. Details of the used instrumental configurations
and exposure times are provided in Table \ref{log}. 
The spectrophotometric standards, Hiltner\,600, GD\,153, and Ross\,640 \citep{1998AA...335L..65H, 2000AA...355L..27H, 2012MNRAS.426.1767P} 
were observed with the R500R+GR, R2500R, and R300R gratings, respectively, on the same night of the scientific observations.
Observations of the standards with R300R and R500R were done with the same grating combined with the Sloan $z$ filter to correct for second-order 
contamination beyond 9200\,\AA~(see procedure in \citealt{2014A&A...568A...6Z}).
Bias frames, continuum lamp flat fields, and Neon\,$+$\,Xenon arc lamps were observed by the observatory staff during the afternoon preceding 
the observations.

We reduced the OSIRIS spectra using routines within {\sc iraf}. We subtracted the raw spectra 
by a median-combined bias and divided by a normalized continuum lamp flat field. From the optimally combined
2D images we extracted the spectra using the {\sc apall} routine and calibrated in wavelength with 
the lines from the combined arc lamp. To improve the detection of the secondary in observations 
with R300R we inserted both components in the slit and used the trace of the primary as a 
reference to combine the individual exposures. Correction of the instrumental response was done 
using the corresponding spectrophotometric standards observed during the same nights. The obtained spectra are displayed in 
Figures \ref{fullspec}, \ref{prim_spec} and \ref{comp_spec_nir}.

\subsection{NOT/ALFOSC Optical Spectroscopy}
For a first estimate of the radial velocity of the primary star, intermediate-resolution optical spectroscopy was acquired for VHS\,1256-1257~A
using the Andalucia Faint Object Spectrograph and Camera (ALFOSC) instrument of 
the 2.5\,m Nordic Optical Telescope (NOT) operating at the Observatorio del Roque 
de los Muchachos, La Palma. The camera uses a E2V back illuminated CCD42-40 chip 
with 2k$\times$2k pixels with a scale of 0.19 arcsec\,pix$^{-1}$, providing a 
field of view of 6.4$\times$6.4~arcmin$^2$. We observed the primary in two different epochs, 
on the nights of 2014 April 22 and May 5. For the instrument setup we chose Grism\#17 
and a 1\farcs3 slit providing a wavelength range of 6330.4--6853.6\,\AA ~and a 
nominal resolution of 0.255\,\AA\,pix$^{-1}$. On the first night one useful exposure 
of 900\,s was obtained under the presence of variable clouds. On the second night four 
exposures of 900\,s were taken, under clear sky conditions. The seeing was similar on 
both nights (0\farcs8--1\farcs1), and hence the final resolution of the spectra was 
also very similar: 1.5\,\AA ~($R$\,$\sim$\,4700). During the same nights using the same instrumental 
configuration, we also observed the star GJ\,388, which is a M4.5V with a precise radial velocity determination of 
$v_r=12.453\pm0.066$ km\,s$^{-1}$ by \citet{2012arXiv1207.6212C}.

Raw data were reduced using routines within the {\sc iraf} environment. Two dimension 
images were bias corrected and flat fielded using continuum lamps normalized using the 
{\sc response} routine. Spectra were optimally extracted using the {\sc apall} routine 
and were wavelength calibrated using HeNe lamps and a cubic spline function fit of the
order of three, providing an rms better than 0.06\,\AA. Final spectra were corrected by the 
instrumental response using the spectrophotometric standard HZ44.

\subsection{VLT/UVES Optical Spectroscopy}
To measure a more precise radial velocity and to investigate in detail the spectral signatures of youth of the primary star, 
like the Li\,{\sc i} line at 670.82~nm we also obtained a high-resolution spectrum of VHS\,1256-1257~A
using the Ultraviolet and Visual Echelle Spectrograph (UVES) mounted on the Kueyen unit of the 
ESO Very Large Telescope. 
UVES is a two-arm crossdispersed echelle spectrograph covering the 
wavelength range 300--500\,nm (blue) and 420--1100\,nm (red), with the possibility to use 
dichroic beam splitters. The instrument is equipped with a single chip in the blue arm and a mosaic 
of two chips in the red arm. The blue CCD is a 2K$\times$4K, 15\,$\mu$m pixel size thinned 
EEV CCD-44. The red CCD mosaic is made of an EEV chip of the same type and the MIT/LL CCID-20 chip.
Each arm has two cross disperser gratings working in first spectral order; the typical order of
separation is 10\arcsec.
We have used the standard dichroic mode setting, which covers the 565--950\,nm spectral range 
in the red arm, with the central wavelength at 760\,nm, and a 373--499\,nm range in the blue arm, 
centered at 437\,nm. 
This configuration, with 1~arcsec slit, provides spectral
resolution of $R$\,$\sim$\,40\,000 (25~mA/pix). Observations were performed in service mode 
under the ESO DDT program 293.C-5014(A) on 2014 July 1, 9 and 10, with clear sky conditions. The total integration time was
9900\,s divided into three exposures of 3300\,s.

The spectra were reduced, extracted, and calibrated using the ESO UVES pipelines under the 
ESO Recipe Flexible Execution Workbench environment \citep[{\it Reflex,}\,][]{2013A&A...559A..96F}.
The reduction steps of the workflow executed by the software include creation of master flat and bias frames,
with the corresponding corrections of science data and detection of the order positions on the detector.
Subsequently, a wavelength calibration solution is obtained from an input arc-lamp frames.
We have used the ThAr arcs acquired in both blue and red arms, during the afternoon preceding the observations.
Instrument characteristics description provides a note of caution in that in the spectral region above 700~nm the ThAr lamp 
has some very bright Argon lines that saturate the CCDs, and the heavily saturated lines may
contain remnants in the following exposures. Here the calibration frames were taken more than four
hours before the night, i.e. more than the typical time after which the remnants vanish.
Then, the spectrum from each order is extracted and merged. Last, the flux-calibration of the science
spectrum is carried out, using the appropriate instrument response curve, in the blue part obtained from 
a standard star observation, and from the instrument master response curve in the red part. 
We have corrected the telluric absorption features in the spectrum using the ESO {\it Molecfit} 
software \citep{2015A&A...576A..77S, 2015A&A...576A..78K}.
The results obtained using the UVES data are described in Section 4.4, with the final spectra presented in
Figures \ref{uvesspec} and \ref{uvesspec_features}.

\begin{table}
\centering
\caption{Measurements and determined physical parameters of VHS\,1256-1257 system\label{measurements}}
\begin{tabular}{l c c}
\hline
\hline
Astrometry   & Primary & Companion \\
\hline
R.A. (J2000)\tablenotemark{a} &   \hspace*{2pt}12$^{\rm h}$56$^{\rm m}$01$^{\rm s}$.922    &   \hspace*{2pt}12$^{\rm h}$56$^{\rm m}$01$^{\rm s}$.586  \\
Decl.  (J2000)\tablenotemark{a} &  $-$12\degr57\arcmin23\farcs990    &  $-$12\degr57\arcmin30\farcs310  \\
2MASS ID                      & J125602.15-125721.7 & J125601.83-125727.6 \\
Separation (arcsec)\tablenotemark{a}  & \multicolumn{2}{c}{8.06\,$\pm$\,0.03} \\
Separation (AU) & \multicolumn{2}{c}{$102\pm9$ } \\ 
Position angle (deg)\tablenotemark{a} & \multicolumn{2}{c}{218.1\,$\pm$\,0.2}  \\
$\mu_{\alpha}\cos\delta$ (mas\,yr$^{-1}$) & \hspace*{-2.2mm} $-281.5\pm5.3$ & \hspace*{-1.3mm}$-275.4\pm5.3$  \\ 
$\mu_{\delta}$ (mas\,yr$^{-1}$) & $-205.5\pm15.2$ & $-198.4\pm15.2$  \\
Parallax $\pi$ (mas) & \multicolumn{2}{c}{$78.8\pm6.4$}  \\
Distance $d$ (pc) & \multicolumn{2}{c}{$12.7\pm1.0$} \\
$v_r$ (km\,s$^{-1}$) & \hspace*{1.4mm}$-1.4\pm4.5$ & ... \\
{\it U} (km\,s$^{-1}$) & \hspace*{1.4mm}$  -9.4\pm2.0$ & ... \\         
{\it V} (km\,s$^{-1}$) & $ -16.4\pm3.0$  &  ...\\
{\it W} (km\,s$^{-1}$) & \hspace*{1.4mm}$  -8.5\pm3.9$ & ... \\
\hline
Photometry & & \\
\hline
$V$ (mag) & $17.759 \pm 0.059$ & ... \\
$I$ (mag) & $13.318 \pm 0.039$ & ... \\
Sloan $i$ (mag) & ... & $22.494 \pm 0.315$ \\
Sloan $z$ (mag) & ... & $20.095 \pm 0.090$ \\
2MASS $J$ (mag)   & $11.018 \pm 0.023$ & $16.662 \pm 0.287$ \\
2MASS $H$ (mag)   & $10.473 \pm 0.023$ & $15.595 \pm 0.209$ \\
2MASS $K_s$ (mag) & $10.044 \pm 0.021$ & $14.568 \pm 0.121$ \\
VHS $Y$ (mag) & $<$11.72\,\tablenotemark{b} & $18.558 \pm 0.051$ \\
VHS $J$ (mag) & $<$11.36\,\tablenotemark{b} & $17.136 \pm 0.020$ \\
VHS $H$ (mag) & $<$11.02\,\tablenotemark{b} & $15.777 \pm 0.015$ \\
VHS $K_s$ (mag)  & $<$10.42\,\tablenotemark{b} & $14.665 \pm 0.010$ \\
{\it WISE W}1 (mag) & $9.880 \pm 0.023$ & $13.6 \pm 0.5$ \\
{\it WISE W}2 (mag) & $9.658 \pm 0.021$ & $12.8 \pm 0.5$ \\
{\it WISE W}3 (mag) & $9.390 \pm 0.044$ & $>$\,11.8 \\
{\it WISE W}4 (mag) & $8.334 \pm 0.410$ & $>$\,8.65 \\
\hline
Spectral Classification  & & \\
\hline
Optical & M$7.0\pm0.5$ & L$8.0\pm2.0$ \\ 
Near-IR $J$ & M$8.0\pm0.5$ & L$8.0\pm1.0$  \\
Near-IR $K$ & M$8.0\pm0.5$ & L$5.0\pm2.0$ \\
H$_2$O index & M$7.6\pm0.4$ & $>$L4.0 \\
H$_2$OD index & $<$L0 & L$5.8\pm0.8$\\
H$_2$O-1 index & M$6.9\pm1.1$ & ... \\
H$_2$O-2 index & M$7.9\pm0.5$ & ... \\
Adopted spectral type & M$7.5\pm0.5$ & L$7.0\pm1.5$ \\
Near-IR gravity class  & INT-G & VL-G \\
\hline
Physical Properties & & \\
\hline
Age (Myr) & \multicolumn{2}{c}{150\,--\,300} \\
$\log(L_{\rm bol}/L_{\odot})$ & $-3.14\pm0.10$ & $-5.05\pm0.22$ \\
Mass ($M_{\rm Jup}$)  & 73$^{+20}_{-15}$ & 11.2$^{+9.7}_{-1.8}$ \\
$T_{\rm eff}$ (K)  & $2620\pm140$ & 880$^{+140}_{-110}$ \\
$\log(g)$  & $5.05\pm0.10$ & 4.25$^{+0.35}_{-0.10}$ \\
\hline
\end{tabular}
 \tablenotetext{1}{ Measured using VHS images, epoch (MJD) = 55743.067635}
 \tablenotetext{2}{ $YJHK_s$ photometry out of VIRCAM linear range}
\end{table}

\begin{figure*}
\centering
\vspace*{4mm}
 \includegraphics[scale=0.99,keepaspectratio=true]{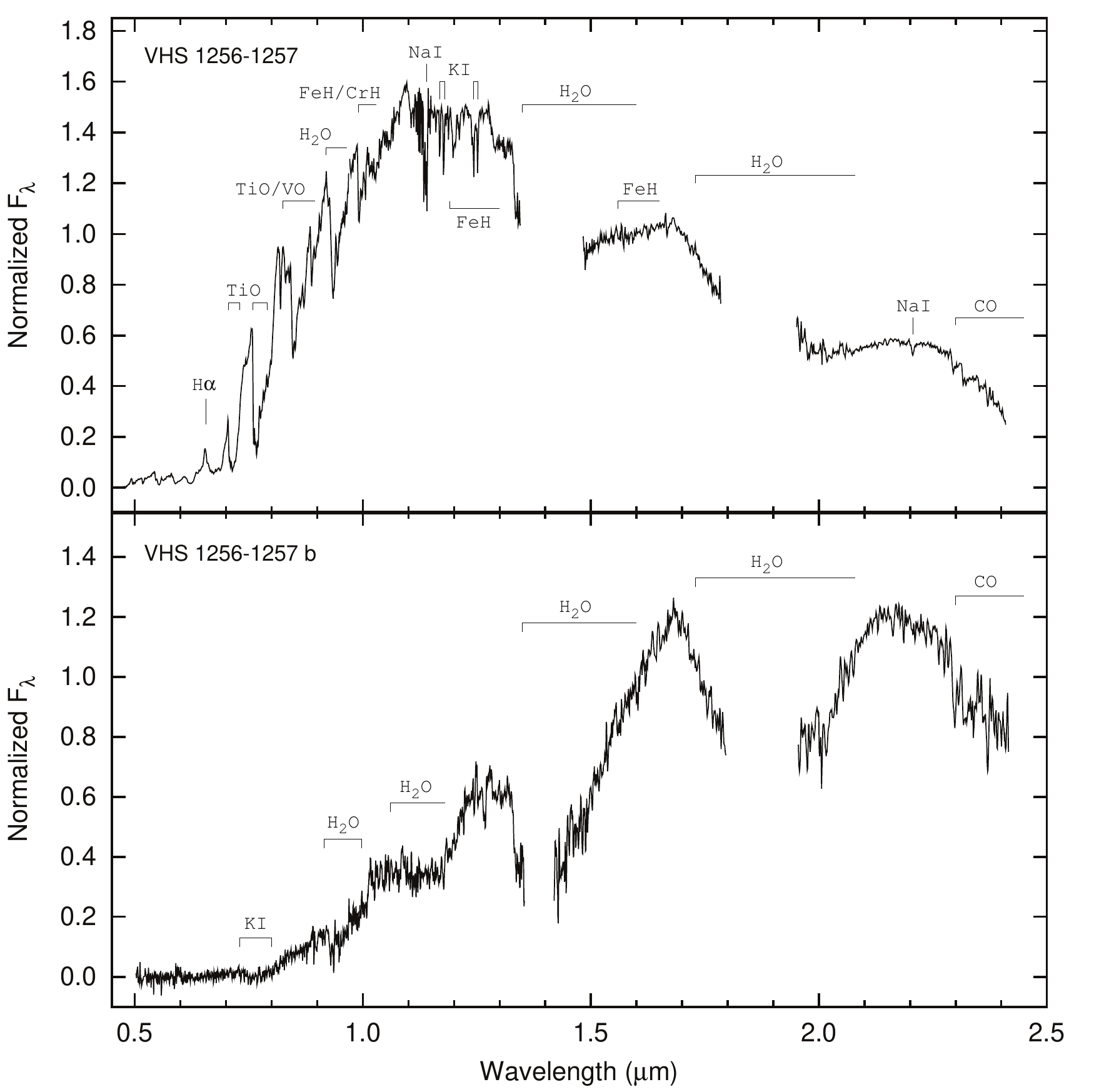}
  \caption{Low-resolution optical (0.5--1.0\,$\mu$m, $R$\,$\sim$\,120--320) and near-infrared 
  (1.0--2.4\,$\mu$m, $R$\,$\sim$\,600) spectra of the primary (top) and companion (bottom) 
  obtained using GTC/OSIRIS and NTT/SofI. The flux is normalized at 1.6\,$\mu$m. The most 
  prominent molecular and atomic features are indicated. Unlike most of old population 
  field L dwarfs, the spectrum of VHS\,1256-1257~b peaks in the $H$ band, which has 
  a triangle-shaped continuum characteristic for young, dusty L dwarfs.}
  \label{fullspec}
\end{figure*}

\subsection{IAC80/CAMELOT Imaging}
On 2014 July 15 and December 19 and 20 we performed imaging observations of the primary using
CCD Camera CAMELOT of the 0.8\,m IAC80 telescope at the Teide Observatory on Tenerife. We aimed to
obtain optical photometry in the $VI$ (Johnson-Bessell's) filters. The images were also used 
in the parallax determination.
The camera contains a E2V 2048$\times$2048 back illuminated chip with 0\farcs304 pixels corresponding to a 
10.4$\times$10.4 arcmin$^2$ field of view. Individual exposures of 300 s were obtained
in the $V$ and $I$ filters. On the night of July 15, right after the scientific target, we observed
two standard star fields at similar airmass, containing eight standard stars used for the 
photometry calibration. Weather conditions were clear/photometric 
with seeing of 1\farcs1--1\farcs3.

\begin{figure*}
 \includegraphics[scale=0.63,keepaspectratio=true]{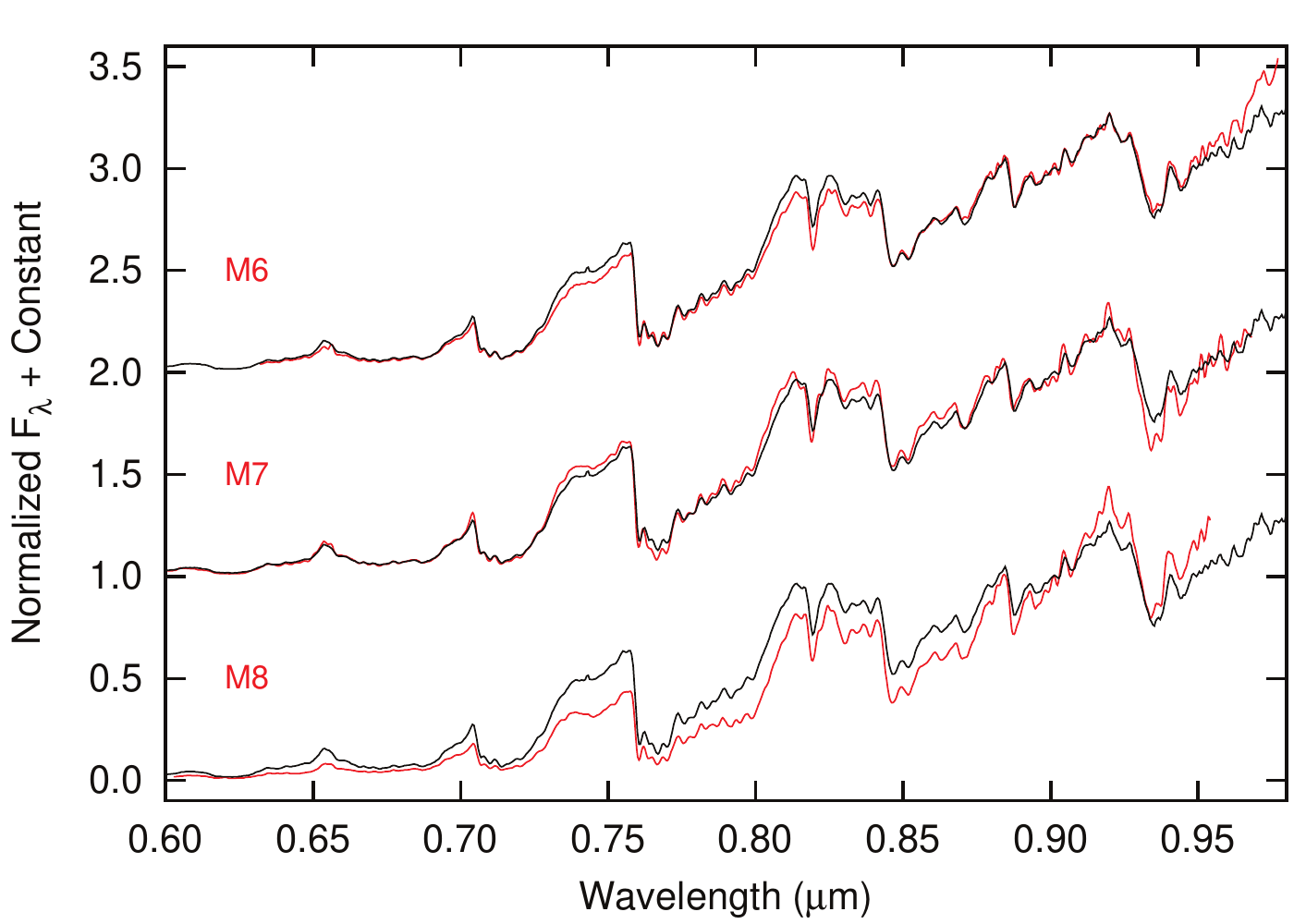}
  \includegraphics[scale=0.63,keepaspectratio=true]{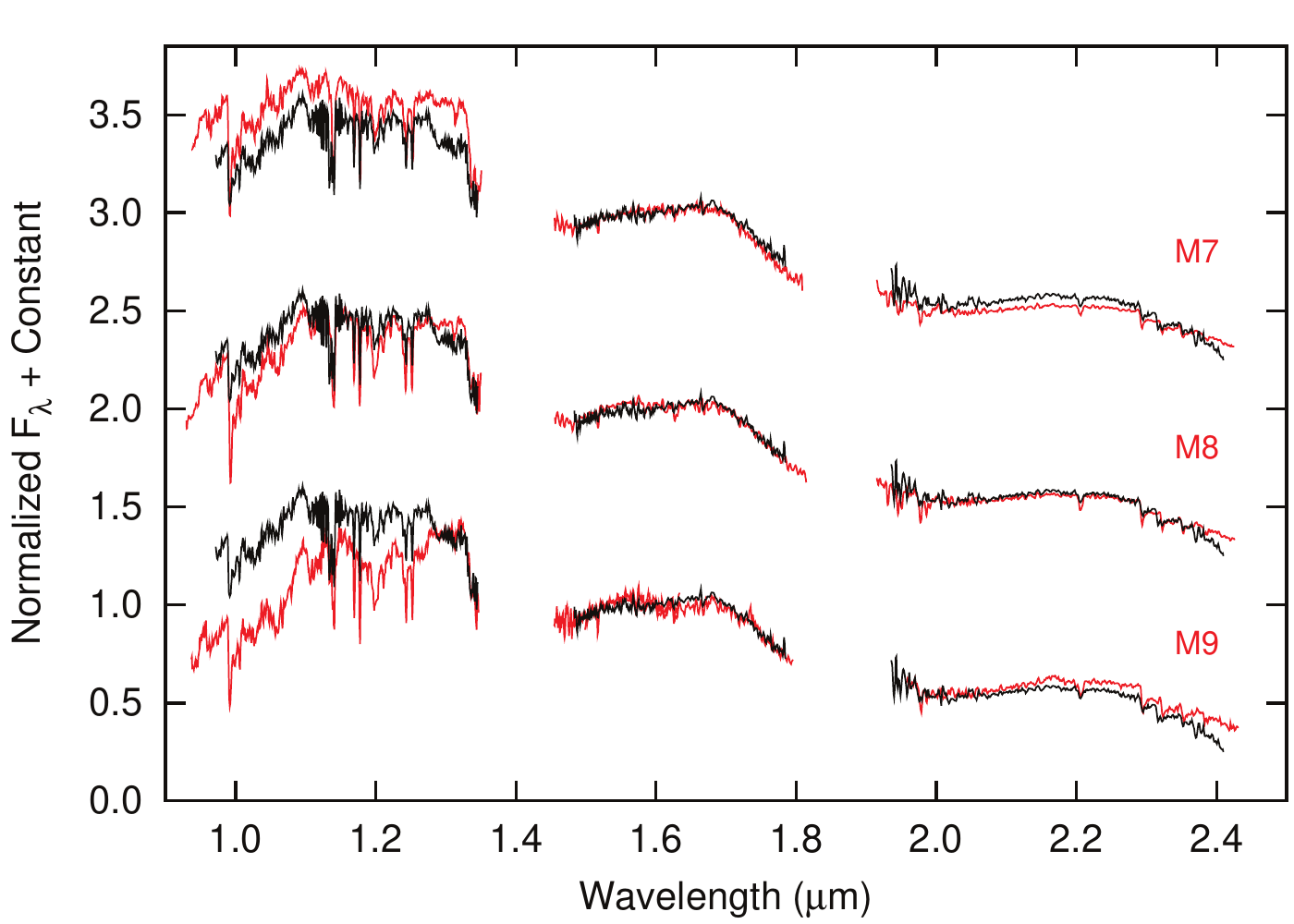}
  \caption{Low-resolution GTC/OSIRIS optical (left) and NTT/SofI near-infrared 
  (right) spectra of the primary VHS~1256-1257 plotted with black line, compared with late-M 
  spectral templates (with labels indicating their type). Comparison spectra were smoothed to 
  match the resolution of our data. The sources and references of used templates are described 
  in Section 4.1.1. Spectra were normalized at 0.9\,$\mu$m at optical and 
  at 1.6\,$\mu$m at near-IR and offset by a constant for display. }
  \label{prim_spec}
\end{figure*}

The data were reduced automatically using a dedicated PyRAF script that includes a standard bias and 
flat-field corrections procedure. We carried out aperture photometry with the {\sc phot} task within the
{\sc iraf} environment. Photometric calibration of the instrumental magnitudes in each filter was 
obtained using eight standard stars \citep{1992AJ....104..340L, 2009AJ....137.4186L} from the two observed 
fields. The uncertainty of the calibration is $\sim$0.04~mag. The measured $VI$ magnitudes of the primary 
and other photometric data of the primary and the companion are listed in Table~\ref{measurements}.

\subsection{WHT/ACAM and LIRIS Imaging}
To complement the photometric information at different wavelengths, we also acquired 
optical images of the pair using the Sloan $i$ and $z$ broad-band filters. The final $z$-band 
image was also employed to constrain the trigonometric parallax (Section 4.2.2).
Observations were done using the ACAM camera at the 4.2-m William Herschel Telescope (WHT) on 2014 July 17.
ACAM uses a 2k$\times$2k pixels EEV CCD detector with 0.25 arcsec\,pix$^{-1}$ scale, providing
a circular field of view with a 8.3~arcmin$^2$ diameter. We used a nine-point dither pattern with
individual exposures of 30\,s. 
On the same observing run on July 18 we have obtained $J$-band observations using the
Long-slit Intermediate Resolution Infrared Spectrograph (LIRIS) spectrograph and imager
on the WHT, for an additional astrometry epoch of the binary used to constrain the parallax measurement.
LIRIS uses a 1k$\times$1k HAWAII detector with a pixel scale of 0.25 arcsec\,pix$^{-1}$, 
yielding a field of view of 4.27$\times$4.27 arcmin$^2$. We used a sequence of five dither 
patterns of 9 positions and 2\,s individual exposures, giving a total exposure time of 
90\,s. Weather conditions were clear during the observations, with average seeing of 
0\farcs85 on the first and 0\farcs66--0\farcs68 on the second night.

We carried out the data reduction of the LIRIS images with the LIRIS data reduction 
package\footnote[7]{\url{www.ing.iac.es/Astronomy/instruments/liris/liris_ql.html}}. The 
procedure consisted of subtracting the sky made from the dithered images, correction for 
flat field, vertical gradient observed on the detector and the geometrical distortion.
ACAM images were reduced with standard procedures using {\sc iraf}. Master sky flat
frames were obtained from science images by combining by the median the individual exposures of the 
dither sequence. Corrected images were aligned and average combined. The aperture photometry
of the companion was obtained using the {\sc phot} task within {\sc iraf}. To calibrate the 
instrumental magnitudes we measured the photometry of stars in a field observed after 
VHS\,1256-1257, under similar weather and airmass conditions. We selected 10 point sources with a 
good signal-to-noise ratio and available measurement in the SDSS Photometric Catalog \citep{2012ApJS..203...21A}.
The obtained {\it i} and {\it z} magnitudes are listed in Table \ref{measurements}.
The primary was saturated in the individual exposures.

\begin{figure}
\centering
\vspace*{1.5mm}
  \includegraphics[scale=0.771,keepaspectratio=true]{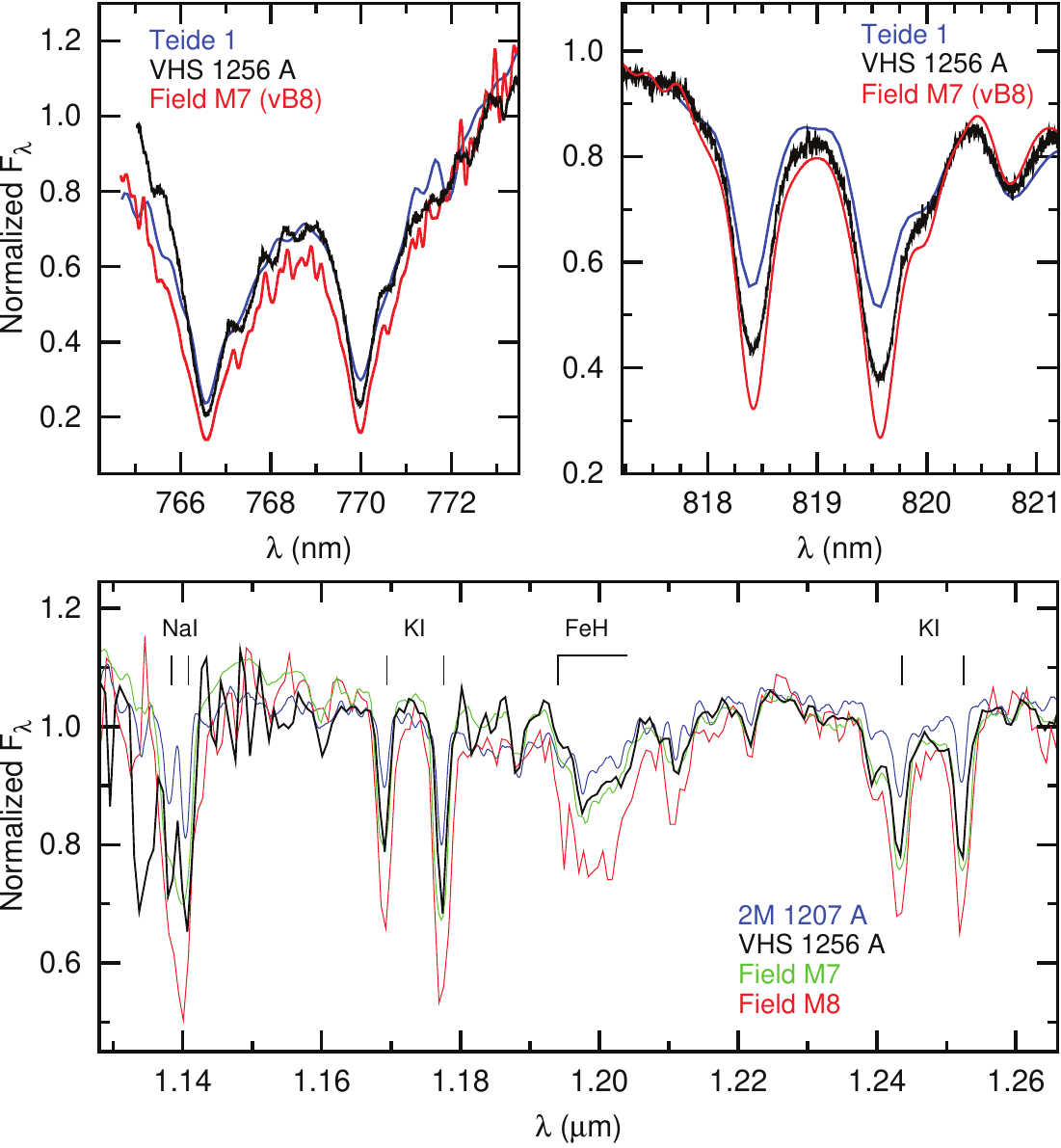}
  \caption{Comparison of the Na\,\textsc i and K\,\textsc i lines of the primary VHS~1256-1257
  in the optical (top panels) and near-IR (bottom panel) with standard field M7--M8 counterparts and with young low-gravity 
  M8-type dwarfs Teide~1 and 2M\,1207\,A. All spectra per panel were conveniently
  normalized at a certain wavelength, and lines were convolved to the same resolution for a proper comparison.  
  These absorption features are slightly less intense than in field objects of similar spectral type.} 
  \label{alkali_lines}
\end{figure}

\subsection{{\it WISE} Data}
The {\it Wide-field Infrared Survey Explorer} \citep[{\it \hspace*{-0.4mm}WISE};][]{2010AJ....140.1868W} conducted a survey of the entire 
sky in the 3.4, 4.6, 12 and 22\,$\mu$m mid-infrared bandpasses (hereafter $W1$, $W2$, 
$W3$ and $W4$). The {\it WISE} All-Sky Source Catalog (see explanatory supplement document by Cutri 
et al. \url{http://wise2.ipac.caltech.edu/docs/release/allsky/expsup/} includes the photometry of 
VHS~1256-1257 in the four bands, but does not detect VHS~1256-1257\,b. The All{\it \hspace*{-0.4mm}WISE} Source 
Catalog contains the deblended photometry of both the primary in the four bands and the secondary 
in the {\it W}1 and {\it W}2, but the later is contaminated by a diffraction spike of the bright star as 
indicated in the confusion flag. In order to obtain the correct values, we performed aperture 
photometry of VHS~1256-1257\,b using {\sc daophot} routine in the {\it W}1 and {\it W}2-band images 
after PSF subtraction of the primary using the scaled PSF of nearby stars in the field as references. 
A small aperture of 5 pixels was used and instrumental magnitudes was transformed into apparent 
magnitudes using the photometry of VHS~1256-1257. The All{\it \hspace*{-0.4mm}WISE} Source Catalog photometry of the 
primary and the derived photometry of the secondary are given in Table \ref{measurements}.

\section{Physical properties of the system}
\subsection{Spectral Types and Spectral Characteristics}
In Figure \ref{fullspec} we plot the merged low-resolution optical (0.5--1.0\,$\mu$m) and near-IR 
(1.0--2.4\,$\mu$m) spectra of each of the two components of VHS\,1256-1257. The spectral 
energy distribution of the primary (top panel) corresponds to that of mid-/late-M dwarfs, 
with the highest flux in the $J$ band, broad water vapor absorption bands cutting out the 
continuum between $JH$ and $HK$ regions and numerous hydride, oxide (FeH, TiO, VO, CO), and 
alkali line (Na\,{\sc i}, K\,{\sc i}) absorption features.
Also, a H$\alpha$ emission line at 656\,nm is detected. The overall appearance of the companion 
spectra (bottom panel) appears to resemble an L-type object, but it is significantly redder 
than field L dwarfs. Unlike a typical L dwarf, its flux peaks at the $H$ band, which has a 
triangular continuum shape, usually interpreted as a hallmark of low surface gravity and youth. In the 
following, we discuss in detail the spectral characteristics, determine the spectral 
types of the binary components, and analyze the gravity-sensitive features in the spectra.

\subsubsection{The Primary}
To determine the spectral types, we used our low-resolution optical and near-IR spectra.
We classified the objects in a qualitative manner through visual comparison of the spectra 
with a set of field dwarf spectral templates, in the optical and near-IR separately.
Subsequently, we used the spectral indices established by \citet{2013ApJ...772...79A} to designate 
spectral types in a quantitative way, and estimate the gravity class. The final spectral 
types were then assigned by averaging the information from both approaches with their 
corresponding uncertainties.

In Figure \ref{prim_spec} we show the optical (left plot) and near-IR (right plot) 
spectra of the primary overplotted with the templates. In the optical, we compared our 
GTC/OSIRIS spectra normalized at 900\,nm with a set of M dwarf spectral templates:
Gl~406 (M6, \citealt{2008AJ....136.1290R}), vB~8 (M7, \citealt{1997ApJ...476..311K}) 
and vB~10 (M8, \citealt{1990ApJ...354L..29H}).
In this wavelength range we find the best match with an M7-type dwarf, while a noticeably poorer fit to templates 
with one subtype difference. Thus, in the optical we assign a spectral type of M7\,$\pm$\,0.5. 
In an analogous manner, we compared the near-IR NTT/SofI spectra normalized at 
1.6\,$\mu$m with M-dwarf spectral templates available in the IRTF Spectral 
Library\footnote[8]{\url{http://irtfweb.ifa.hawaii.edu/~spex/IRTF_Spectral_Library/}} maintained by Michael Cushing \citep{2009ApJS..185..289R}. 
On the right-hand side panel of Figure \ref{prim_spec} we include a 
comparison of the primary spectra with vB~8 (M7, \citealt{2005ApJ...623.1115C, 
2009ApJS..185..289R}), LP 412-31 (M8, \citealt{2009ApJS..185..289R}) all smoothed to match 
resolution of $R$\,$\sim$\,600 and the M9 standard LHS~2065 \citep{1991ApJS...77..417K} which we 
observed using SofI with the same configuration as for VHS~1256-1257. We found the best consistency 
with the M8 standard. Both M7- and M9-type standards provided 
significantly worse matches, especially in the $J$ band. We adopt a near-IR spectral 
type of M$8\pm0.5$, one subtype later than the one designated from the optical.

\begin{table}
\begin{center}
\scriptsize
\caption{Pseudo-equivalent widths (pEW) of lines and doublets measured in the VLT/UVES, 
NTT/SofI and NOT/ALFOSC spectra of the primary VHS\,1256-1257.\label{pew}}
\begin{tabular*}{\columnwidth}{l c c c}
\hline
\hline
Line  & Date (UT) & $\lambda$ (nm) & pEW (\AA)\\
\hline
H$_\eta$ & 2014 Jul 01 & 383.5 & $-6.06\pm0.91$  \\
H$_\zeta$ & 2014 Jul 01 & 388.9 &$-7.94\pm1.23$   \\
Ca\,{\sc ii} K & 2014 Jul 01 & 393.4 & $-35.63\pm4.42$ \\
Ca\,{\sc ii} H+H$_\epsilon$ & 2014 Jul 01 & 396.8\,+\,397.0 & $-28.44\pm3.79$  \\
H$_\delta$ & 2014 Jul 01 & 410.2 & $-9.49\pm1.78$  \\
H$_\gamma$ & 2014 Jul 01 & 434.1 & $-8.43\pm1.87$  \\
H$_\beta$ & 2014 Jul 01 & 486.1  & $-7.49\pm0.31$  \\
H$_\alpha$ & 2014 Apr 22 & 656.3 & $-3.8\pm0.5$ \\
H$_\alpha$ & 2014 May 04 & 656.3 & $-4.0\pm0.5$ \\
H$_\alpha$ & 2014 Jul 01 & 656.3 & $-4.04\pm0.05$  \\
H$_\alpha$ & 2014 Jul 09 & 656.3 & $-3.74\pm0.19$  \\
H$_\alpha$ & 2014 Jul 10& 656.3 & $-2.80\pm0.09$  \\
Li\,{\sc i} & 2014 Jul 01 & 670.8 & $<0.03$\\
K\,{\sc i} doublet & 2014 Jul 01 & 766.5\,/\,769.9 & 9.35$\pm$0.17\,/\,5.94$\pm$0.16 \\
Na\,{\sc i} doublet & 2014 Jul 01 & 818.3\,/\,819.8 & 1.80$\pm$0.06\,/\,2.88$\pm$0.06 \\
Na\,{\sc i}        &  2014 Mar 12 & 1139  &   $8.2\pm0.8$ \\
K\,{\sc i} doublet & 2014 Mar 12 & 1169\,/\,1177 & 3.0$\pm$0.6\,/\,4.2$\pm$0.7 \\
K\,{\sc i} doublet & 2014 Mar 12 & 1244\,/\,1253 & 4.0$\pm$0.7\,/\,3.4$\pm$0.6 \\
\hline
\end{tabular*}
\end{center}
\end{table}

As mentioned above, in parallel to visual comparison of the overall spectral morphology with 
the standards, we calculated the spectral indices defined by \citet{2013ApJ...772...79A}. 
We used the H$_2$O, H$_2$OD, H$_2$O-1, and H$_2$O-2 indexes, optimized to avoid spectral features 
dependent on the surface gravity of the object and to have a well-established correlation 
with optical spectral types. The H$_2$O index, which measures the slope of the steam absorption 
at the blue end of the $H$ band ($\sim$1.50--1.57\,$\mu$m), yields a spectral type of M$7.6\pm0.4$.
From H$_2$O-1 and H$_2$O-2 indexes we obtained spectral types of M$6.9\pm1.1$ and 
M$7.9\pm0.5$, respectively. Both from the comparison with templates and from the calculated 
indices, we found the typing to be consistent within the uncertainties. As the final spectral 
type we adopt an M$7.5\pm0.5$, obtained as the mean of all the types inferred from visual 
comparison and from spectral indexes, weighted by their corresponding uncertainties.

\begin{figure*}
  \includegraphics[scale=0.62,keepaspectratio=true]{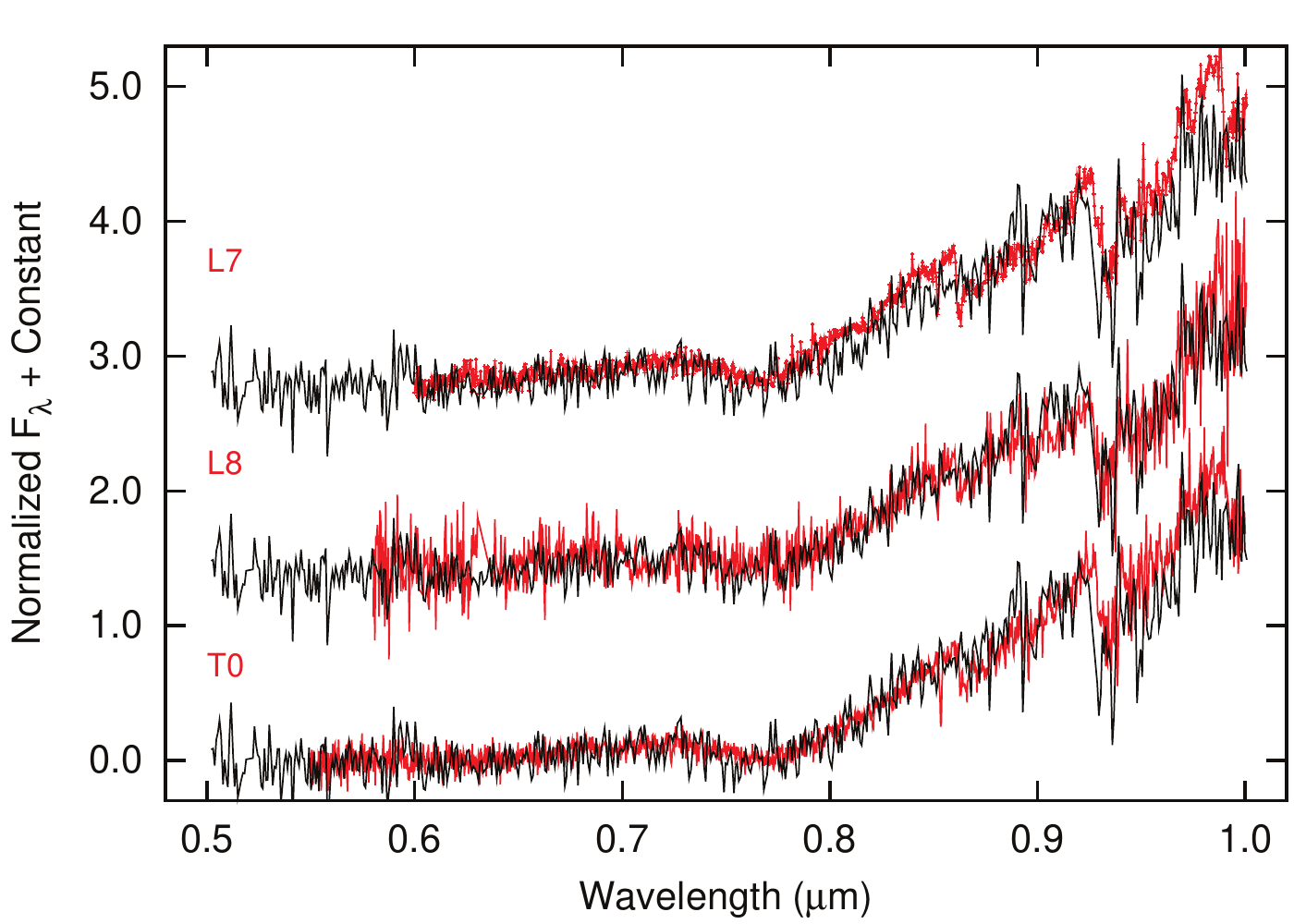}
  \includegraphics[scale=0.62,keepaspectratio=true]{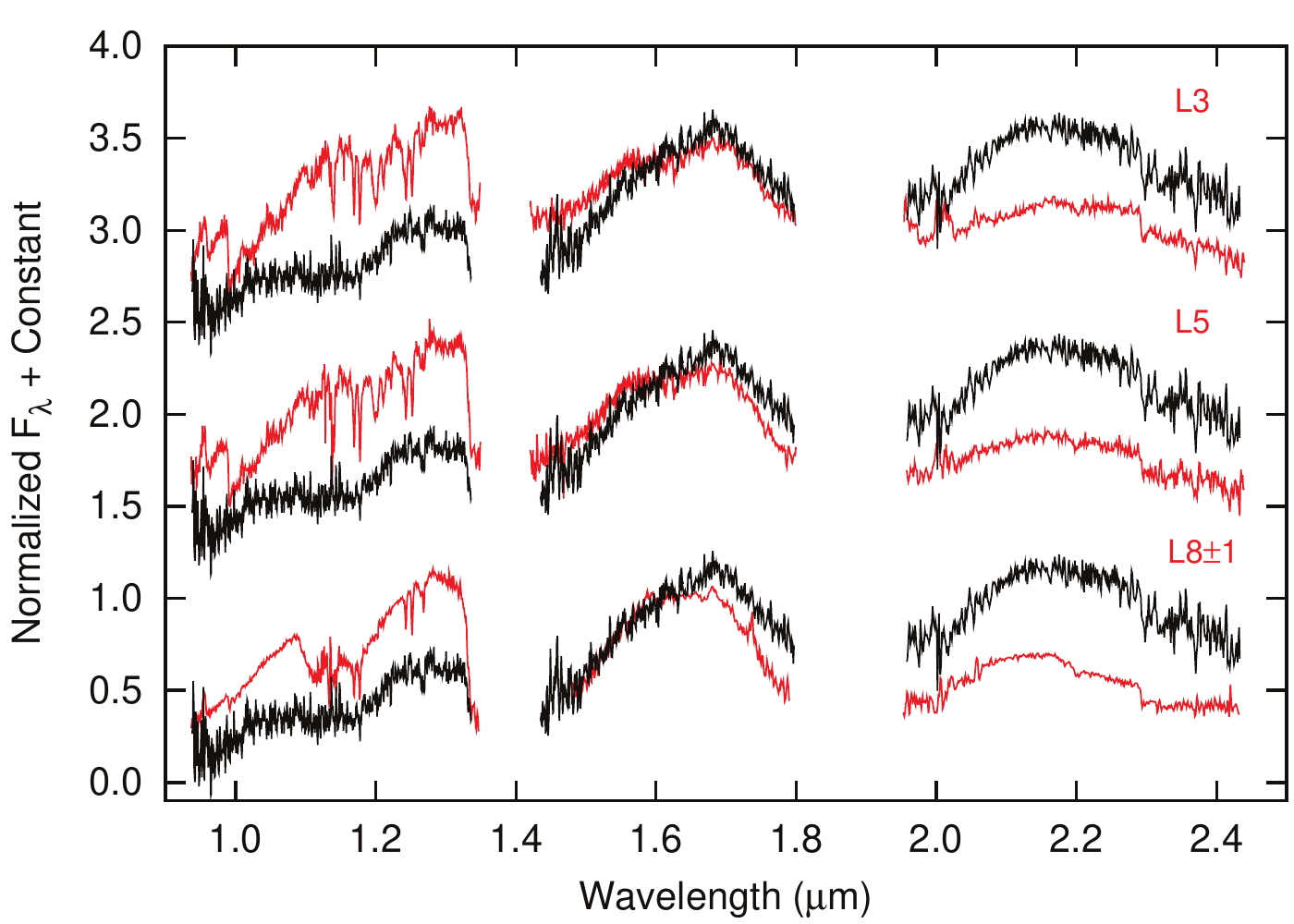}
  \caption{Low-resolution GTC/OSIRIS optical (left) and NTT/SofI near-infrared 
  (right) spectra of the companion VHS\,1256-1257~b plotted with black line,
  compared with high-gravity field dwarf templates. The sources and references of 
  used templates are described in Section 4.1.2. Spectra were normalized at 
  0.9\,$\mu$m at optical and at 1.6\,$\mu$m at near-IR and offset by a constant 
  for display.}  
  \label{comp_spec_nir}
\end{figure*}

\begin{figure*}
\centering
  \includegraphics[scale=0.83,keepaspectratio=true]{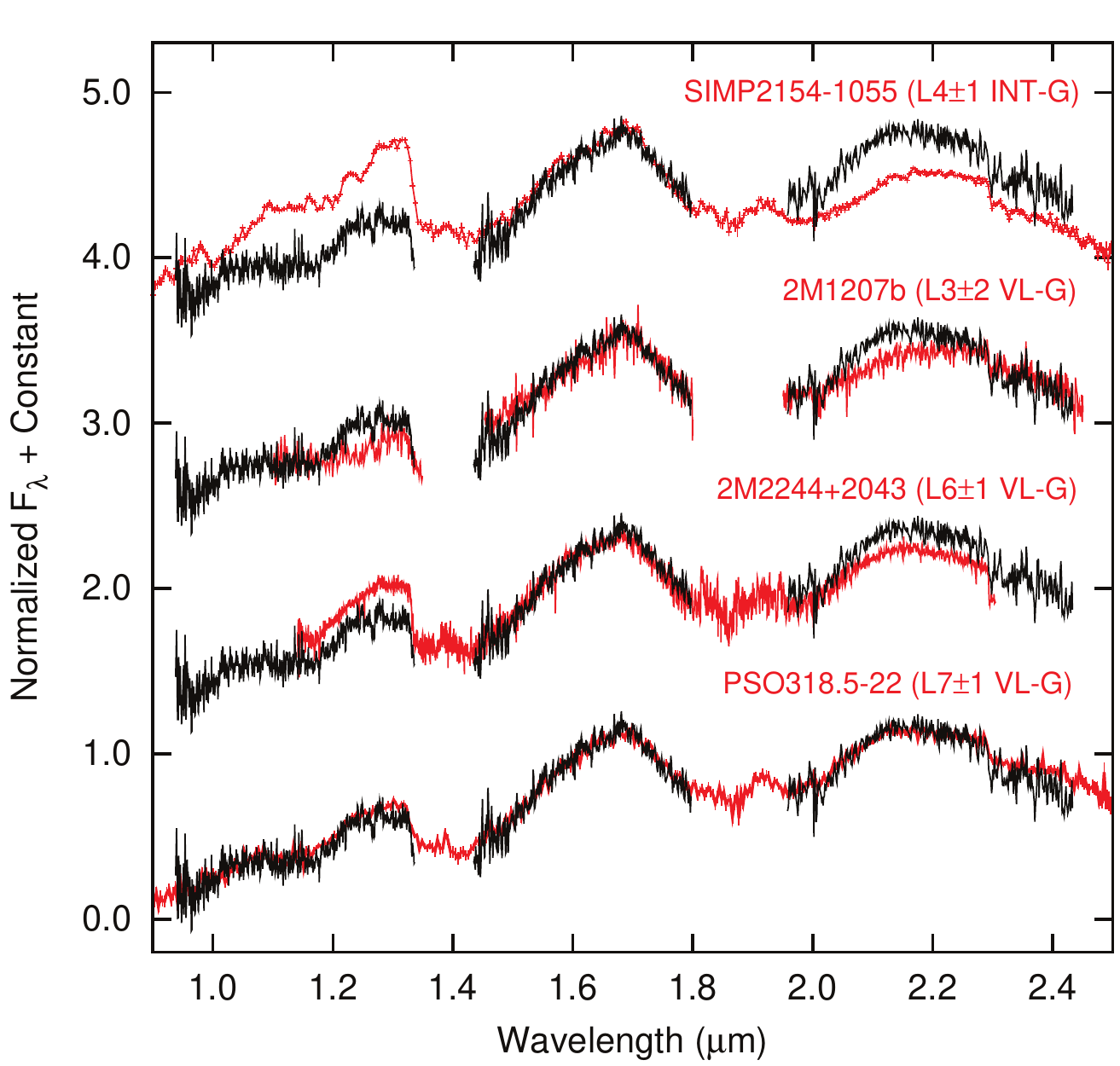}
  \caption{SofI/NTT low-resolution near-infrared spectra (0.9--2.4~$\mu$m, $R$\,$\sim$\,600) 
  of the companion VHS\,1256-1257~b (plotted in black) compared with an L$4\pm1$ type dwarf 
  SIMP2154-1055 classified as intermediate surface gravity object (INT-G) and with a set of young, 
  very low-gravity (VL-G) L dwarfs (red lines). Spectra were normalized at 1.6 $\mu$m and 
  shifted by a constant for display. The spectrum of VHS\,1256-1257~b is best matched with spectrum 
  of the free-floating planet PSO\,J318.5-22 classified as L$7\pm1$ type.}
  \label{comp_spec_young}
\end{figure*}

To find out if the primary has characteristics of low surface gravity, 
we inspect the gravity-sensitive spectral features like the Na\,{\sc i} 
and K\,{\sc i} lines. In Fig. \ref{alkali_lines} we present a 
comparison of K\,{\textsc i} and Na\,{\sc i} doublets in the optical 
at around 770 and 819 nm, respectively (top panels of the figure), and in 
the near-IR in a region of the $J$ band (bottom panel). In the optical, 
we plot in black the UVES spectrum of the primary VHS\,1256-1257, in red 
the standard M7 dwarf vB~8 (using UVES spectra retrieved from the ESO Archive) 
and in blue the M8 member of the Pleiades, Teide~1 \citep{1996ApJ...469L..53R}. 
In the near-IR, our NTT/SofI spectrum of VHS\,1256-1257 (black line) is 
overplotted with the field dwarfs vB~8 (M7), LEHPM 2-436 (M8) observed by us with the
same instrumental configuration, and with an M8 at 5--10\,Myr, 2M~1207\,A 
from \citet{2014A&A...562A.127B}. The spectra were smoothed to match the 
resolution of VHS\,1256-1256 data and, in the optical, convolved with a 
Gaussian function to adjust the FWHMs of the lines.

We find these absorption features of the primary to be slightly less intense 
than in the field counterparts, for example, the Na\,{\sc i} lines
at 818.3 and 819.8 nm have pEW of $1.80\pm0.06$
and $2.88\pm0.06$ \AA , respectively. In comparison, from the vB~8 spectra
we measure pEW of $1.98\pm0.07$ and $3.40\pm0.06$ \AA. The same lines in the 
120 Myr old M8-type counterpart Teide~1 are in turn less intense with pEW of
$1.66\pm0.09$ and $2.66\pm0.10$ \AA. 
Because the Na\,{\sc i} subordinate lines are well known to be highly
sensitive to atmospheric pressure at cool temperatures, this indicates that the
primary likely has an intermediate surface gravity that lies
between that of Pleiades and field M8-type dwarfs. We note, however,
that all these comparisons are based on the assigned spectral types and assume
that both targets and reference sources have the same metallicity. Furthermore,
the intensity of the alkali lines also depend on the effective temperature (or 
spectral type). Given a half-subtype uncertainty in our typing of
VHS\,1256-1257\,A, we found that the surface gravity of this object may 
be consistent with field objects considering an earlier type.

In Table \ref{pew} we provide the pEW of Na\,{\textsc i}, K\,{\textsc i}, 
and other lines measured in the optical and near-IR spectra of the primary.
In order to quantify the gravity class we have calculated four spectral indices:
FeH$_z$, VO$_z$, K\,\textsc i$_J$, and $H$-continuum, identified to be related to gravity-dependent features and optimized to be employed 
with low-resolution near-IR spectra (\citealt{2013ApJ...772...79A} and references therein). 
Based on these indicators we considered
field gravity (FLD-G), intermediate gravity (INT-G), or very low gravity (VL-G) of an object 
following the classification scheme of \citet{2013ApJ...772...79A}.
For FeH$_z$, which measures the depth of the FeH feature at 0.99 $\mu$m, we obtained 
a value of $1.13\pm0.08$. For the VO$_z$ index we assign a score of ``n'' since it is not gravity sensitive in this 
range of spectral types. For the K\,\textsc i$_J$ index, which measures the depth of potassium 
absorption at 1.244 and 1.253 $\mu$m, we got a value of $1.08\pm0.07$
and for the $H$-cont index, which measures the slope of the $H$ band 
we got $0.98\pm0.24$. For an M7.5 spectral type these index values correspond to 
a surface gravity intermediate between that of counterparts classified
as very low gravity and of the old, field population.

\subsubsection{The Secondary}
Our GTC/OSIRIS optical spectrum of the companion (left plot of Fig. \ref{comp_spec_nir}) is of 
modest signal-to-noise ratio, and apart from the general spectral energy distribution only few spectral 
features are noticeable, like the strong K\,{\sc i} resonance doublet and water vapor at 0.92~$\mu$m. 
The lack of oxide features (TiO, VO) at the resolution of the visible spectrum suggest a type later than mid-L.
On the left panel of Fig. \ref{comp_spec_nir} we show the spectra of the companion (black line) 
normalized at 0.9 $\,\mu$m and overplotted with objects of spectral types L7 (2MASS J21522609+0937575, 
\citealt{2008AJ....136.1290R}), L8 (2MASS J03400942-6724051, \citealt{2008AJ....136.1290R}) and T0 
(2MASS J04234858-0414035, \citealt{2003AJ....126.2421C}). The template spectra were retrieved from 
the Ultracool RIZzo Spectral Library. Comparison with templates of field dwarfs yields a spectral 
type of L$8\pm$2. 

From the comparison of the NTT/SofI near-IR spectrum of the companion VHS\,1256-1257~b to 
L-type objects with well-determined spectral types, we find that the overall 1.0--2.4~$\mu$m spectral energy distribution 
does not match any early to late high-gravity field L dwarf. In the right-hand side plot of Fig.~\ref{comp_spec_nir} 
we show a comparison of the companion near-IR spectra normalized at 1.6~$\mu$m (black line) with 
an L3-type object DENIS-P J1058.7-1548 \citep{2010ApJ...710.1142B}, L5 SDSS J053951.99-005902.0 \citep{2005ApJ...623.1115C}
and L$8\pm1$ {\it WISE} 1049-5319\,A \citep{2013ApJ...767L...1L}. 
The spectra of these L dwarf templates were obtained by us using SofI instrument with the same configuration as for VHS\,1256--1257
and were reduced in the same way.
In the $J$ band, VHS\,1256--1257~b is relatively
less luminous than normal field L dwarfs and contrarily is significantly brighter in the $K$ band. 
Its $H$-band continuum unlike the field L dwarfs has a distinctive triangular shape, interpreted 
as a signature of low surface gravity and youth.
To measure the near-IR spectral type, we tried to find best matching field dwarf standards from the IRTF library, 
at the $J$ (1.07--1.40~$\mu$m) and $K$ (1.90--2.20~$\mu$m) windows separately, following the approach of \citet{2013ApJ...772...79A}.
In this wavelength regions the shape of spectra continuum is expected to have a lower dependence 
on gravity. We estimate, in the $J$ and $K$ bands, respectively, a spectral type of L$8\pm1$ and L$5\pm2$. 
In the $K$ band we assign an uncertainty of two subtypes because several standards provide similarly 
good fits.

Along with visual comparison, we calculated the H$_2$OD index, well defined up to the range of 
late-L dwarfs. The other indexes are not valid for spectral types later than L5. The H$_2$OD 
index value give a spectral type of L$5.8\pm0.8$. 
In analogous manner as for the primary, from a mean of the types inferred through visual comparison 
and from the H$_2$OD spectral index, weighted by the corresponding errors, we obtained an L7.0 type, 
which we adopt here as the spectral type of the companion, with an uncertainty of 1.5 subtype.
Among the gravity-sensitive spectral indices, only the $H$-continuum index is well defined up to L7 type objects.
For VHS\,1256-1257\,b $H$-cont indicates a very low gravity (VL-G).
However, as noted by \cite{2013ApJ...772...79A}, the shape of the $H$-band continuum is not the most 
reliable indicator, and can provide rather only a hint of low gravity. It should be used in  
combination with other gravity-sensitive indices, but for the spectral type in question none
of them have yet been defined. 
The K\,{\sc i} lines at 1.17 $\mu$m (pEW\,$\le$\,5\,\AA) and at 1.25 $\mu$m (pEW\,$\le$\,3\,\AA) appear weaker in 
the spectrum of the secondary than in the field mid- and late-L dwarfs (see Fig.~\ref{comp_spec_nir}).
For example, in the spectrum of the L8 dwarf {\it WISE}~1049-5319\,A obtained with the same instrument and 
setup we measured a pEW of the K\,{\sc i} lines of 5.7$\pm$0.8 and 6.8$\pm$0.7~\AA~at 1.169 and 1.177~$\mu$m 
and 3.4$\pm$0.4 and 6.1$\pm$0.4~\AA~at 1.244 and 1.253~$\mu$m, respectively. 
This also indicates that the secondary has a lower gravity compared to field counterparts \citep{2004ApJ...600.1020M}.

\begin{sidewaystable*}
\vspace*{7cm}
\begin{center}
\caption{Images used for astrometric measurements. \label{par_observations}} 
\begin{tabular}{lclccccc cccc}
\hline  \hline
\multicolumn{1}{l}{Epoch} & 
\multicolumn{1}{c}{JD-2400000} & 
\multicolumn{1}{l}{Tel/Instrument} & 
\multicolumn{1}{c}{Filter} & 
\multicolumn{1}{l}{Pixel} & 
\multicolumn{1}{c}{Seeing} & 
\multicolumn{1}{c}{$\rho$} &
\multicolumn{1}{c}{$\theta$} &
\multicolumn{2}{c}{Primary} &
\multicolumn{2}{c}{Secondary} \\

\multicolumn{1}{l}{(UT)} & 
\multicolumn{1}{c}{} & 
\multicolumn{1}{l}{} & 
\multicolumn{1}{c}{} & 
\multicolumn{1}{l}{(mas)} & 
\multicolumn{1}{c}{($''$)} & 
\multicolumn{1}{c}{($''$)} & 
\multicolumn{1}{c}{($\deg$)} &
\multicolumn{1}{c}{$\Delta$R.A.\,(mas)} & 
\multicolumn{1}{c}{$\Delta$Decl.\,(mas)} &
\multicolumn{1}{c}{$\Delta$R.A.\,(mas)} & 
\multicolumn{1}{c}{$\Delta$Decl.\,(mas)} \\   

\hline
1999 Mar 1\tablenotemark{a} & 51238.7109 & 2MASS           & $J$ & 999 & 3.1 & 7.35$\pm$0.21 & 217.1$\pm$1.5 & 3217$\pm$103 & 2305$\pm$153 & 3868$\pm$115 & 2830$\pm$178 \\
2011 Jul 1\tablenotemark{b} & 55743.5676 & VISTA/VIRCAM    & $Y$ & 339 & 1.0 & 8.06$\pm$0.03 & 218.1$\pm$0.2 &  0  &  0  &  0  &  0 \\
2014 Mar 13\tablenotemark{c} & 56729.7276 & NTT/SofI        & $J$ & 288 & 0.8 &  ...          & ...             & $-$633.9$\pm$24.8 & $-$604.9$\pm$27.6 & $-$660.3$\pm$22.2 & $-$566.8$\pm$30.6 \\
2014 Apr 27 & 56774.7212                  & NTT/SofI        & $J$ & 288 & 1.3 & 8.08$\pm$0.03 & 218.0$\pm$0.2 & $-$727.3$\pm$6.6 & $-$590.6$\pm$6.0 & $-$722.2$\pm$17.6 & $-$621.9$\pm$11.5   \\
2014 Apr 27 & 56774.7224                  & NTT/SofI        & $H$ & 288 & 1.0 & 8.05$\pm$0.03 & 217.9$\pm$0.2 & $-$756.0$\pm$25.3 & $-$607.8$\pm$21.0 & $-$728.7$\pm$25.3 & $-$610.2$\pm$21.0 \\       
2014 Jul 15\tablenotemark{d} & 56854.4126 & IAC80/CAMELOT   & $I$ & 304 & 1.1 &  ...          & ...             & $-$871.5$\pm$26.7 & $-$645.4$\pm$25.2  & ... & ... \\
2014 Jul 15\tablenotemark{d} & 56854.4139 & IAC80/CAMELOT   & $I$ & 304 & 1.1 &  ...          & ...             & $-$834.3$\pm$14.9 & $-$682.3$\pm$11.6  & ... & ... \\
2014 Jul 17\tablenotemark{e} & 56856.3903 & WHT/ACAM        & $z$ & 254 & 0.9 &  ...          & ...             & ... & ... & $-$833.0$\pm$17.5 & $-$647.2$\pm$17.5 \\
2014 Jul 18 & 56857.3642                  & WHT/LIRIS       & $J$ & 250 & 0.7 & 8.06$\pm$0.04 & 218.0$\pm$0.3 & $-$855.1$\pm$16.8 & $-$578.6$\pm$15.5   & $-$842.6$\pm$16.7 & $-$583.6$\pm$15.5 \\
2014 Dec 19\tablenotemark{d} & 57010.7372 & IAC80/CAMELOT   & $I$ & 304 & 1.5 &  ...         & ...              & $-$835.3$\pm$28.0 & $-$738.8$\pm$31.0 & ... & ... \\
2014 Dec 19\tablenotemark{d} & 57010.7447 & IAC80/CAMELOT   & $I$ & 304 & 1.5 &  ...         & ...              & $-$799.8$\pm$30.0 & $-$762.2$\pm$32.4 & ... & ... \\
2014 Dec 20\tablenotemark{d} & 57011.7568 & IAC80/CAMELOT   & $I$ & 304 & 1.2 &  ...         & ...              & $-$825.8$\pm$35.7 & $-$744.1$\pm$38.8 & ... & ... \\
\hline
\end{tabular}
\end{center}
\tablenotetext{1}{ Excluded from $\pi$ determination.}
\tablenotetext{2}{ Reference epoch.}
\tablenotetext{3}{ Primary is slightly saturated.}
\tablenotetext{4}{ Astrometry is derived for primary only (secondary is too faint).}
\tablenotetext{5}{ Astrometry is derived for secondary only (primary is highly saturated).}
\end{sidewaystable*}

In addition to the comparison with field L-type templates, we have also compared the near-IR 
spectra of VHS\,1256-1257~b with a set of known young low-gravity mid- and late-L dwarfs. 
In Figure \ref{comp_spec_young} we overplot the companion spectra with SIMP~J2154-1055, 
classified as an L$4\pm1$\,$\beta$ with intermediate surface gravity 
\citep{2014ApJ...792L..17G}, a $\sim$5 $M_{\rm Jup}$ planetary mass companion 
2MASS~1207-3932\,b at 8 Myr \citep{2005A&A...438L..25C, 2010A&A...517A..76P}, 
2MASS J22443167+2043433 classified as very low-gravity L$6\pm1$ type \citep{2003ApJ...596..561M} 
and with the free-floating planetary mass object PSO J318.5-22 \citep{2013ApJ...777L..20L}. 
In general, the spectrum of the companion fits much better to the spectra of young, very 
low-gravity L dwarfs than to the spectra of field objects. We find the best match with 
PSO J318.5-22 which is classified as an L$7\pm1$, thus providing a further support of our 
spectral type classification. PSO J318.5-22 is a 6.5~$M_{\rm Jup}$ object, possible member of 
the 12 Myr moving group $\beta$ Pictoris. Because of this strong resemblance, we 
state that the companion can be assigned to the very low-gravity (VL-G) class.

\subsection{Distance Estimates}
\subsubsection{Spectrophotometric Distance}
Having classified the spectral type of the primary, we used the $JHK_s$ photometry to 
estimate the distance of the system. We used the 2MASS catalog photometry, since at 
these magnitudes the VHS measurements start to get beyond the linear regime of the 
detector and could be uncertain. 
We employed the mean absolute magnitudes as a function of spectral type determined by
\citet{2012ApJS..201...19D} from a set including 8 M7.0, 9 M7.5, and 11 M8.0 type dwarfs
with precise parallactic distance determination.
Considering an M$7.5\pm0.5$ type of the primary, and assuming that it is
a single object, we inferred spectroscopic distances of 12.0$^{+5.8}_{-3.6}$, 
12.4$^{+4.7}_{-3.5}$ and 12.2$^{+4.3}_{-3.3}$ pc, using $J$, $H$, and $K_s$ photometry, 
respectively. Taking the average of three bands, we estimate a distance of 12.2$^{+5.0}_{-3.5}$ pc. 
The quoted errors account for uncertainties of the photometry, spectral type, and the intrinsic dispersion in 
absolute magnitudes for field dwarfs at a given spectral type.

In this case however, the estimation that employs the M dwarfs found in the field should be
considered with caution, since the field objects have ages typically of more than 1 Gyr,
and for VHS\,1256-1257 we expect an age below 300 Myr (see Section 4.5). From the studies of young star clusters, it is
known that the M dwarfs younger than 400--625 Myr can be overluminous with respect to their field
counterparts \citep{2006A&A...458..805B, 2008MNRAS.385.1771J, 2014arXiv1410.2383Z}. 
For that reason we obtained an alternative estimation using near-IR photometry 
of M7--M8 dwarf members of the Pleiades cluster, compiled by \cite{2010A&A...519A..93B}.
The Pleiades has an age of $120\pm10$ Myr \citep{1996ApJ...458..600B, 1998ApJ...499L..61M, 
1998ApJ...499L.199S} and is located at a distance of 133.5 pc \citep{2005AJ....129.1616S, 
2014Sci...345.1029M}. Averaging the estimates obtained in $J$, $H$, and $K_s$ yields a
distance of 14.9$^{+3.3}_{-3.2}$ pc for VHS\,1256-1257 if this binary had the age of the Pleiades.

\subsubsection{Trigonometric Parallax}

To derive the trigonometric parallax and proper motion of the pair we used an area 
of $5' \times 5'$ around the target and the images listed in Table~\ref{par_observations}, 
except for the 2MASS. 
We took advantage of the availability of these images for a first determination of a trigonometric distance to 
the system. In Table~\ref{par_observations} we provide the observing epoch (including the 
Julian Date), telescope, and imaging instrument, filter, the average pixel projection onto 
the sky, and the mean seeing of the images. All data were properly reduced following 
standard steps for the optical and near-IR wavelengths as explained in Section 3. We selected the VISTA/VIRCAM $Y$-band
observations of 2011 July as the fundamental frame to which all other images are compared. 
Using the {\sc daofind} command within {\sc iraf} we identified all sources with photon 
peaks with detection above 6 $\sigma$, where $\sigma$ stands for the noise of the background, 
and FWHM resembles that of unresolved objects (i.e., extended sources were mostly avoided). 
In addition, we ensured that the detected sources lied within the linear regime of the detectors 
response (with the only exception of the primary component of the binary). The number of 
sources identified per image in common with the reference frame ranged from 25 to $\sim$40. 
The centroids of detected objects were computed by estimating the $x$ and $y$ pixel positions 
of the best fitting one-dimensional Gaussian functions in each axis; typical associated errors 
are about 3\%--5\%~of a pixel or better. 

\begin{figure}   
\center
\includegraphics[scale=0.8]{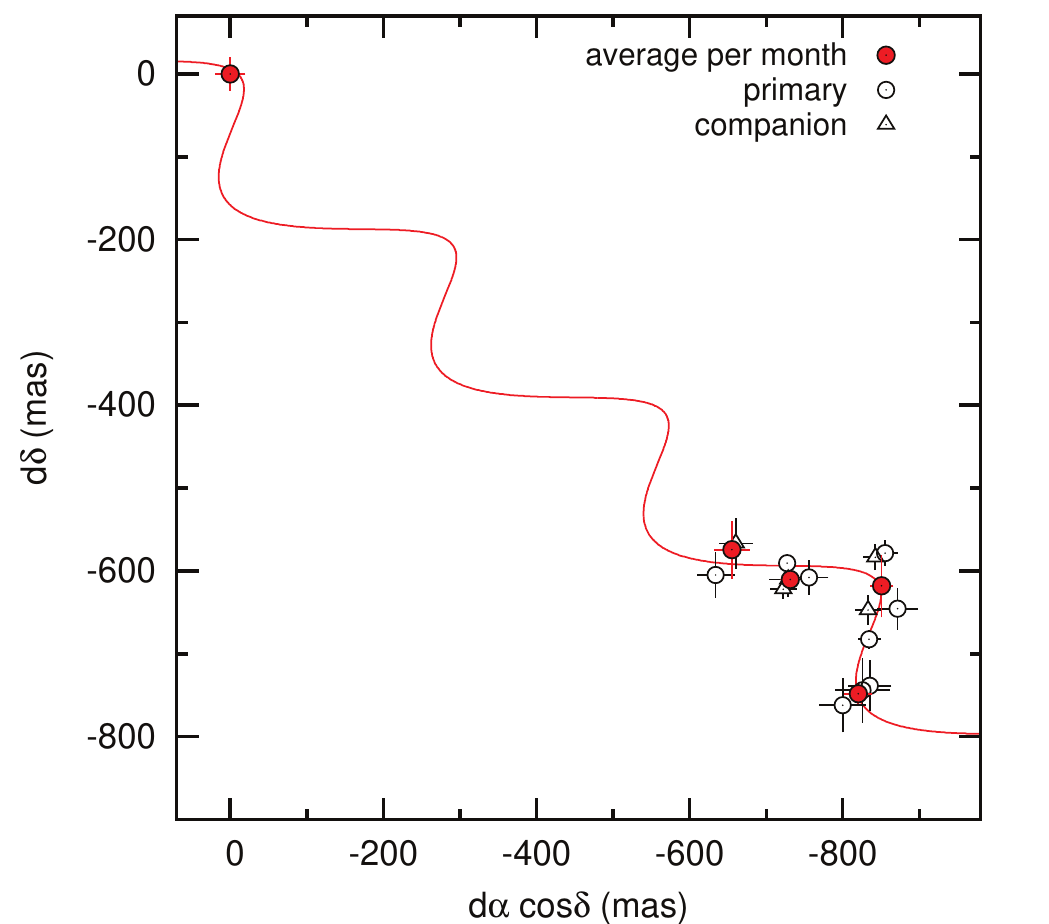}
\caption{Relative apparent trajectory of the components of VHS\,1256-1257 in right ascension ($\alpha$) and 
declination ($\delta$) coordinates for the time interval 2011 July (reference epoch) through 2014 December. 
Each component is plotted separately as indicated in the figure legend, and the averaged values per 
observing month are shown with the red solid circles. The solid line indicates the best model fit. North is up and east is to the left. 
  \label{trigpar}}
\end{figure}

Pixel coordinates were transformed between different epochs using the {\sc geomap} routine 
within {\sc iraf}, which applied a polynomial of the third and fourth order in $x$ and $y$ and 
computed linear terms and distortions terms separately. The linear term included an $x$ and 
$y$ shift and an $x$ and $y$ scale factor, a rotation, and a skew. The distortion surface 
term consisted of a polynomial fit to the residuals of the linear term. The $(x,y)$ astrometric 
transformation between observing epochs and the reference epoch was an iterative step, which 
included the rejection of objects deviating by more than 1.5--2 $\sigma$, where $\sigma$ 
corresponds to the dispersion of the transformation. Typical coordinates transformation 
dispersions ranged from $\pm$0.05 to $\pm$0.09 pixels; this is typically $\pm$15--25~mas. 
The centroids of the two components were calculated with an uncertainty similar to that of the astrometric
reference sources. 
Errors associated with the astrometry are dominated by the errors in the coordinate transformations 
for each axis (R.A., decl.), which include the dispersion of all reference sources.
We therefore assigned the dispersion of the coordinates transformation solutions
to the errors associated with the differential astrometry of our targets listed in Table 4.
We did not apply any correction for differential chromatic refraction since all of our 
data were taken at red wavelengths, where the corrections are smaller than the quoted 
astrometric uncertainty. The relative $(dx,dy)$ astrometry (conveniently derotated) was 
converted into d$\alpha$\,cos\,$\delta$ and d$\delta$ using the corresponding plate scales. 
The apparent trajectory of each component, which depends on proper motion 
($\mu_\alpha$, $\mu_\delta$) and parallax ($\pi$), is shown in Figure~\ref{trigpar} and can 
be modeled with the following equations:
\begin{equation}
\label{eq1}
{\rm d}\alpha = \mu_\alpha ~ (t-t_o) + \pi ~ (f^\alpha_t - f^\alpha_o)
\end{equation}
\begin{equation}
\label{eq2}
{\rm d}\delta = \mu_\delta ~ (t-t_o) + \pi ~ (f^\delta_t - f^\delta_o),
\end{equation}
where $t$ stands for time, the subscript $o$ indicates the reference epoch, and $f^\alpha$ and 
$f^\delta$ refer to the parallax factors in right ascension ($\alpha$) and declination ($\delta$), 
respectively. In our study, all the astrometric quantities are given in mas and the times $t$ and 
$t_o$ are measured in Julian Days. The parallax factors were computed by following the equations 
given in \citet{1985spas.book.....G} and obtaining the Earth barycenter from the DE405 
Ephemeris\footnote[9]{\url{http://ssd.jpl.nasa.gov}}. We applied the least-squares fitting method to the 
set of Equations \ref{eq1} and~\ref{eq2} to derive the parallax and the proper motion of the system. 
The two components were fit simultaneously. The best-fit solution yielded 
$\mu_\alpha\,{\rm cos}\,\delta$\,=\,$-277 \pm 5$~mas\,yr$^{-1}$, 
$\mu_\delta$\,=\,$-203 \pm 12.5$~mas\,yr$^{-1}$ 
giving a total proper motion of $\mu$\,=\,$344\pm13$~mas\,yr$^{-1}$,
position angle of the proper motion vector of $233.7\degr\pm2.5\degr$, and a relative parallax  
$\pi$\,=\,77.79\,$\pm$\,6.4~mas. It is depicted with a solid curve in Figure~\ref{trigpar}. 
The amplitude and position angle of the proper motion we determined is in agreement with the measurement given by \cite{2007A&A...468..163D}.
From the fitting performed for each pair member individually we obtained proper motion of
$\mu_\alpha\,{\rm cos}\,\delta$\,=\,$-281.5\pm5.3$~mas\,yr$^{-1}$,
$\mu_\delta$\,=\,$-205.5\pm15.2$~mas\,yr$^{-1}$ for the primary and 
$\mu_\alpha\,{\rm cos}\,\delta$\,=\,$-275.4\pm5.3$~mas\,yr$^{-1}$,
$\mu_\delta$\,=\,$-198.4\pm15.2$~mas\,yr$^{-1}$ for the companion.
The above values are listed in Table \ref{measurements}. 
We have also tested the {\it WISE}, 2MASS, and available earlier epoch data (DENIS, USNO-B1.0, DSS) in the proper 
motion and parallax determination, but despite the longer baseline they introduced larger 
errors to the fitted values. Hence, for the determination of the parallax and proper motion 
we excluded these data, given that their associated uncertainties are at least one order 
of magnitude higher than those of the recent data.
We remark that given the time coverage of our data, which span 3.0447 yr, the proper motion is more 
precisely determined than the parallax. Additional images, in particular taken on months that were not covered yet, are required 
for decreasing the uncertainty in the parallax measurement.

We did not correct our proper motion for the motions of the stars used as a reference because we 
assumed that these (small) motions are randomly orientated with a negligible net effect. However, 
we did apply the correction for converting the relative parallax into the absolute parallax. This 
takes into account that the reference stars are located at finite distances, which diminish part 
of the true parallax of our targets. We followed a procedure similar to the one described in 
\citet{2012ApJ...752...56F} and \citet{2014A&A...568A...6Z}. Using the 2MASS colors 
\citep{2006AJ....131.1163S} of the reference 
objects, we obtained their photometric distances by assuming that all of them are main sequence 
stars. We adopted the color--bolometric correction--spectral type relations given in \citet{1966ARA&A...4..193J} 
for BAFGK stars and in \citet{1993ApJ...402..643K} for late-K and M stars. The defined relations are valid for 
colors in the interval $J-K_s$\,=\,$-$0.2 to 1.53~mag. We adopted the mode of the distribution of 
reference objects distances as the correction to be added to the relative parallax that comes 
directly from our fit to obtain the absolute parallax. The absolute parallax is $\pi$\,=\,78.79\,$\pm$\,6.4~mas, 
which translates into a distance of $12.7\pm1.0$ pc, consistent with the value from the spectrophotometric
estimates using field counterparts. This suggests that the primary component is not an equal-mass binary.

\subsection{Angular Separations and Orbital Motion}
On the available images where both components were detected and the primary did not saturate we measured 
the projected angular separation ($\rho$) and position angle ($\theta$) of the companion using the centroid 
positions of the two sources in each of the images transformed into the VHS $\alpha$, $\delta$ coordinates 
as explained above. The determined values are given in Table \ref{par_observations}. The angular separations 
and position angles of VHS\,1256-1257 in the most recent epoch images over the last three years are consistent 
within $\delta \rho<30$\,mas and $\delta \theta<0.04$\,deg, respectively. Both components share the same proper 
motion, and given the low probability to find a very red young L dwarf (only a few tens found in the whole 
sky area) at this very short angular separation of 8\farcs06, we conclude that both objects are gravitationally 
related. These results also indicate that we do not detect an orbital motion higher than 10~mas\,yr$^{-1}$ 
(at the 1$\sigma$ level).

We expect an orbital period of about 3900~yr from the estimated masses
of the primary and companion (Section 4.7) and a physical projected separation of $\sim$102\,AU. 
Assuming a circular orbit, this 
implies that the displacement caused by orbital motion would be of 4--13~mas\,yr$^{-1}$ depending on the
orientation of the orbit, which is consistent with our results. The angular separation of the pair in the 2MASS images 
is lower, although consistent within 3$\sigma$. These differences (0\farcs6\,$\pm$\,0\farcs2) 
can not be explained by the orbital motion of the companion, since we expect no more than $\sim$\,13~mas\,yr$^{-1}$
for a face-on circular orbit at the physical separation found, and may be attributed to systematic errors 
in the determination of the centroids in the 2MASS data probably due to small separation of the components and low resolution of the images.

\subsection{Kinematics}
To measure the heliocentric radial velocity of VHS~1256-1257 we employed the high spectral resolution 
UVES data (mean Modified Julian Date, MJD\,=\,56838.3530) and the cross-correlation method against the 
M6V star GJ\,406, 
which has a known, constant radial velocity of $v_h$\,=\,19.5\,$\pm$\,0.1~km\,s$^{-1}$ \citep{2002ApJS..141..503N}
and a small projected rotational velocity of $v$\,sin\,$i$ $\le$ 3 km\,s$^{-1}$ 
\citep{2010ApJ...710..924R}. GJ\,406 was also observed with the VLT/UVES instrument (spectral resolution 
of $\sim$45,000) on 2009 March 14 (MJD\,=\,54904.0839). We downloaded the reduced UVES spectrum 
of GJ\,406 from the ESO data archive (program 082.D-0953). Although GJ\,406 was observed with a non-standard 
instrumental setup in terms of wavelength coverage \citep[see][]{2011A&A...534A.133F}, the overlap between 
the M6V-type star and VHS~1256-1257 data is significant and secures a reliable velocity determination. 
The telluric contribution was removed from the data of GJ\,406 in the same manner as we did for 
VHS~1256-1257 (see Section 3.4). 

The telluric-free UVES spectra of VHS~1256-1257 and GJ\,406 were cross-correlated using the task 
{\sc fxcor} within {\sc iraf}. We fit a Gaussian function to the peak of the cross-correlation distribution. 
The resulting relative displacement was corrected for the lunar, diurnal, and annual velocities to obtain 
the heliocentric radial velocity of VHS~1256-1257. All UVES spectra were calibrated in wavelength using 
ThAr arc lines taken some time after the science observations (Section 3.4). The typical stability of 
UVES along the spectral axis is on the order of a pixel over the course of a night; therefore, deviations 
of a km\,s$^{-1}$ can be expected. To correct our velocity determination for this effect, we employed the 
absorption lines from the Earth's atmosphere present in the data, which provide a reliable reference frame 
within $\pm$90 m\,s$^{-1}$ \citep[e.g.,][]{1973MNRAS.162..255G}. We used the UVES spectra of VHS~1256-1257 and 
GJ\,406 before subtraction of the sharp telluric lines to determine the instrumental shifts of the wavelength 
solution for each object. We found them to be $+$0.62\,$\pm$\,0.20 km\,s$^{-1}$ (VHS~1256-1257) and 
$-$0.14\,$\pm$\,0.20 km\,s$^{-1}$ (VHS~1256-1257 versus GJ\,406). The final corrected heliocentric radial 
velocity of VHS~1256-1257 is $v_h$\,=\,$-$1.4\,$\pm$\,5.0~km\,s$^{-1}$, where the error bar accounts for the 
uncertainties due to the cross-correlation procedure, the corrections of the wavelength reference frames, 
and the error associated with the velocity of the M6V standard star.

We checked this heliocentric velocity using the line centroid method, which is independent of any reference star. 
By measuring the centroids of alkali and other metallic lines in the three UVES spectra of VHS~1256-1257 and 
considering the correction for the wavelength reference frame and lunar, diurnal, and annual velocities as 
previously discussed, we found $v_h$\,=\,$+$4.1\,$\pm$\,4.0~km\,s$^{-1}$. Albeit having a much lower spectral 
resolution, the NOT/ALFOSC spectra of VHS~1256-1257 also yielded radial velocities that agree with the 
UVES result at the 1$\sigma$ level (where $\sigma$ stands for the velocity uncertainty): 
$v_h$\,=\,$+$0.1\,$\pm$\,7.0~km\,s$^{-1}$ (MJD\,=\,56769.0443) and $+$14.1\,$\pm$\,10.0~km\,s$^{-1}$ (MJD\,=\,56781.9229). 
The ALFOSC velocities were derived using the cross-correlation technique; the M4.5V star GJ\,388 
($v_h$\,=\,$+$12.6215~km\,s$^{-1}$, \cite{2013A&A...549A.109B}; $v_h$\,=\,$+$12.453~km\,s$^{-1}$, \citealt{2012arXiv1207.6212C}) 
acted as the velocity reference source because its spectra were acquired with exactly the same instrumental 
configuration and observing dates as our target (Section 3.3). GJ\,388 was found to be a radial velocity variable star
by \citet{2013A&A...549A.109B}. However, the amplitude of the variations is 0.25 km\,s$^{-1}$, which is about $\ge$30 times 
smaller than the velocity precision we were able to achieve with the ALFOSC data. No obvious velocity change larger 
than $\sim$10~km\,s$^{-1}$ is observed in VHS~1256-1257 over the 69 day interval of spectroscopic observations.

\begin{figure}
\vspace*{1.1mm}
 \includegraphics[scale=0.71,keepaspectratio=true]{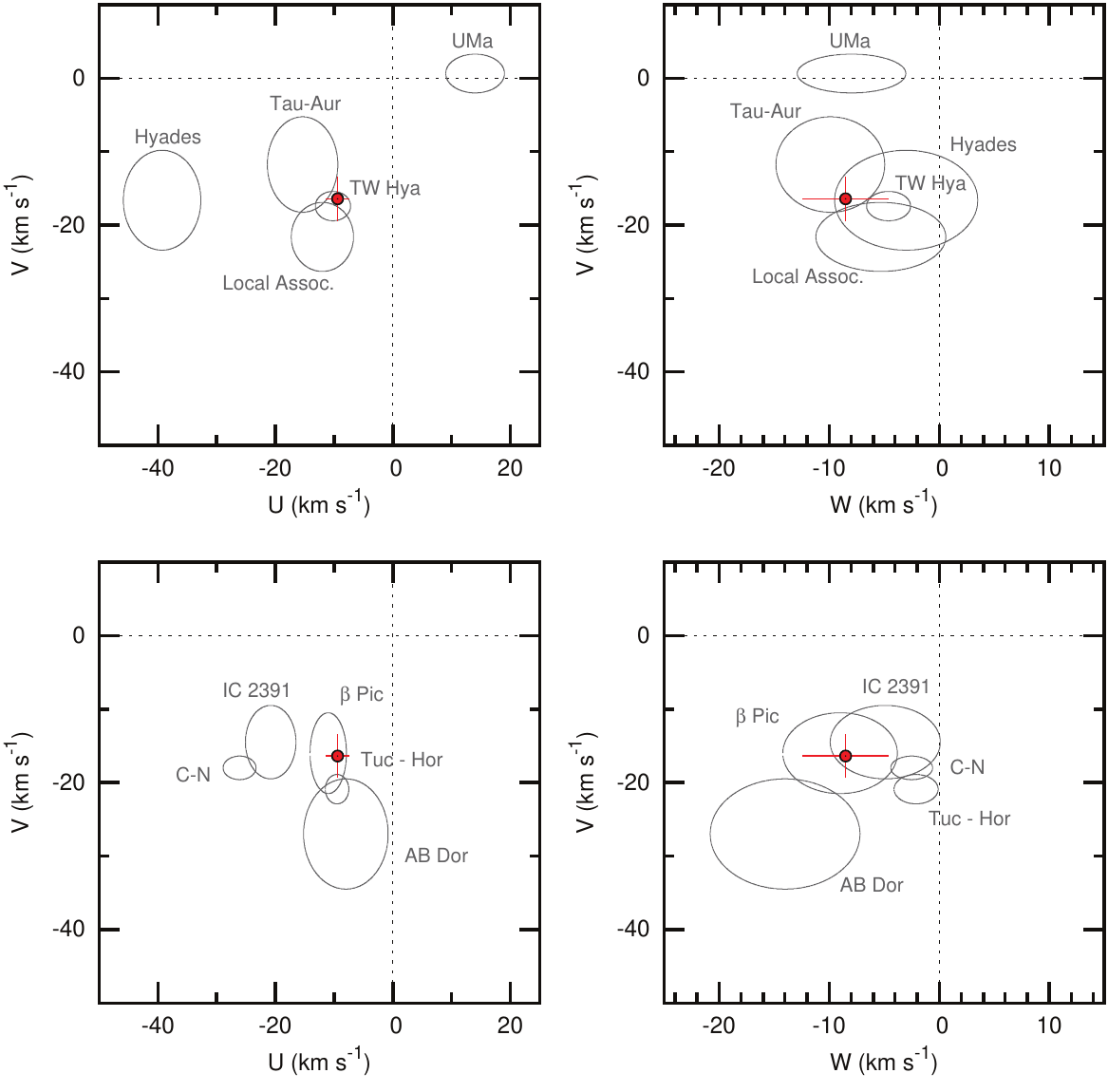}
  \caption{Galactic space velocities of VHS\,1256-1257 (red dots) with overplotted 
  ellipsoids of known young star associations and moving groups. 
  Errors incorporate uncertainties in the proper motion, parallactic distance, and 
  radial velocity. Galactocentric {\it U} velocity is positive toward the Galactic center.}
  \label{uvw}
\end{figure}

\begin{figure*}
 \includegraphics[scale=0.69,keepaspectratio=true]{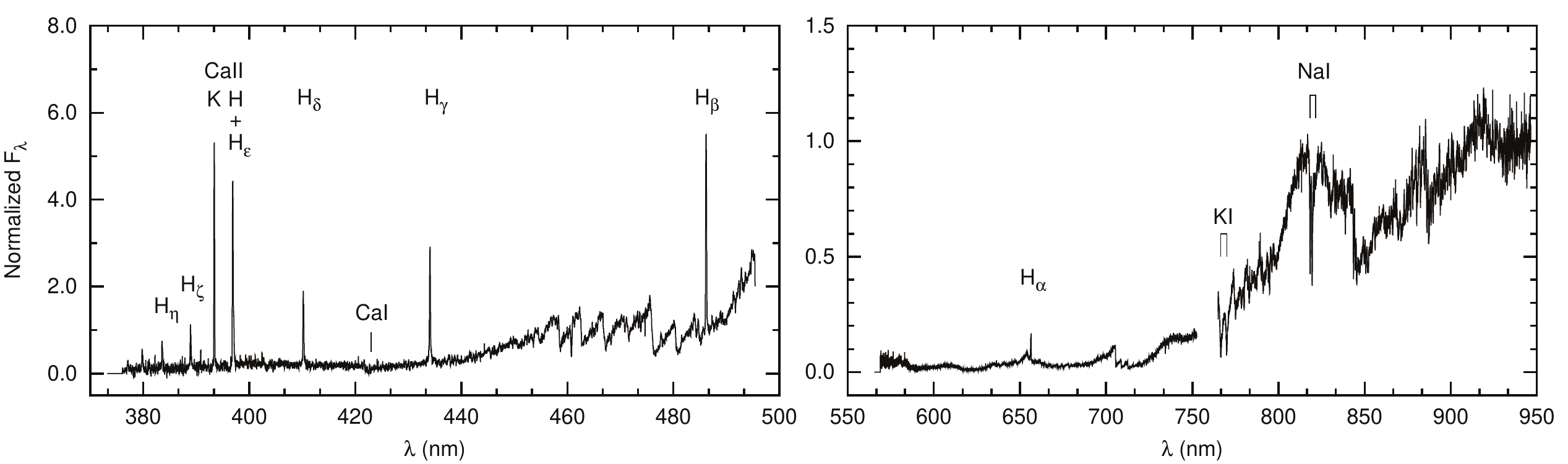}
  \caption{VLT/UVES spectra of the primary VHS\,1256-1257 covering 373--495 and 565--950 nm wavelength range at 
  a resolution of $R$\,$\sim$\,40\,000 (25 m\AA\,pix$^{-1}$). The blue part of the spectrum (left panel) was smoothed for the display 
  by a factor of 33, degrading the resolution to $R$\,$\sim$\,10\,000. 
  The spectrum has been corrected for telluric absorptions using the ESO {\it Molecfit} software. 
  Certain features indicating strong magnetic activity in the chromosphere are visible: 
  the Balmer series emission lines from H$_{\alpha}$ up to H$_{\eta}$, singly ionized calcium 
  H and K lines at 396.85 and 393.37 nm, respectively.}
  \label{uvesspec}
\end{figure*}

Having the measurements of proper motion, parallax, and radial velocity we can determine the three 
components of the Galactic space velocity, {\it U, V,} and {\it W}. We applied the formulas presented 
by \citet{1987AJ.....93..864J} to derive the {\it UVW} velocities listed in Table \ref{measurements} 
and displayed in Figure \ref{uvw}. The uncertainties associated to all three Galactic velocities come from 
the proper motion, parallax, and radial velocity error bars. Figure \ref{uvw} also illustrates the ellipsoids 
corresponding to well-known young stellar moving groups of the solar neighborhood (data compiled 
from \citealt{2004ARA&A..42..685Z} and \citealt{2008hsf2.book..757T}). 

As seen in Figure \ref{uvw}, the space velocities of VHS\,1256-1257 fits with the $\beta$\,Pic Group 
and is consistent within error bars with the galactic velocities of Taurus-Auriga, 
TW\,Hya, and the Local Association groups. The stars in $\beta$\,Pic, Taurus-Auriga, 
and TW\,Hya are younger than 20\,Myr, and this age is incompatible with the non-detection 
of lithium in the atmosphere of the primary (see next section). Hence the 
kinematic and spectroscopic properties of VHS\,1256-1257 is only compatible with the 
membership to the Local Association Group, a coherent kinematic stream of young stars 
(all below 300 Myr, \citealt{1992AJ....104.2141E}) with constituent clusters and 
associations such as the Pleiades (120 Myr), $\alpha$ Persei (50--80 Myr), 
and IC 2602 ($\sim$70 Myr). Membership in stellar moving groups is commonly used to constrain or 
confirm the ages especially of young ($<1$ Gyr) objects. We discuss in more detail the age of 
VHS\,1256-1257 system in the following section.

\subsection{Activity, Lithium and Age}

\begin{figure}
 \includegraphics[scale=0.9,keepaspectratio=true]{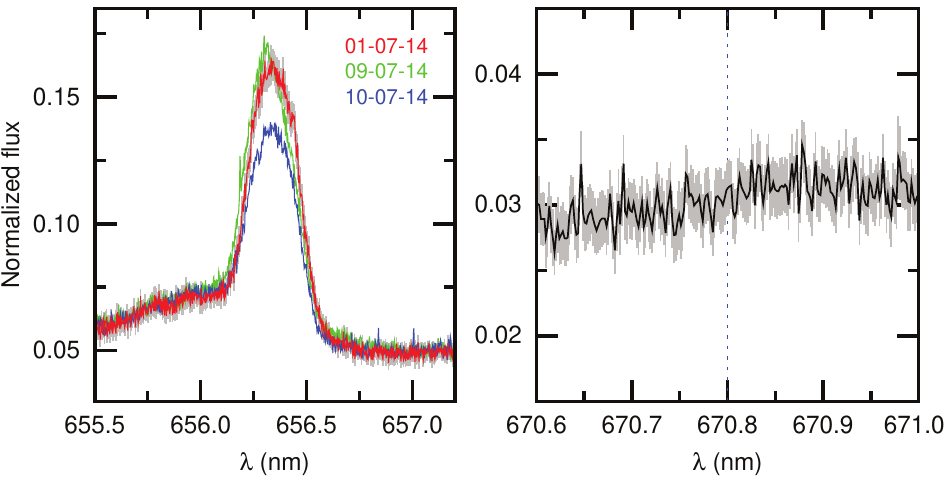}
  \caption{Close-up of regions of the VLT/UVES spectrum of the primary VHS\,1256-1257, showing, 
  in the left panel the H$\alpha$ emission line at 656.3 nm and in the right panel the non-detection 
  of lithium absorption at 670.8 nm. Flux uncertainties are plotted in gray.}
  \label{uvesspec_features}
\end{figure}

The high-resolution optical spectra of the primary (Fig.\,\ref{uvesspec},~\ref{uvesspec_features}) 
reveals emission features recognized as indicators of  
chromospheric activity in very low-mass stars. We detect the Balmer series emission lines of 
hydrogen, from H$_\alpha$ up to H$_\eta$, and single ionized calcium Ca\,{\sc ii} H and K 
emission lines at 396.8 and 393.4 nm. Measurements of the lines pEW are given in Table~\ref{pew}. 
The pEW of H$_\alpha$ does not show a strong variability in timescales of days and months, but show
small variations of the order of 1--2\,\AA~along the observations, as can be seen in Fig.\,\ref{uvesspec_features}
and Table~\ref{pew}.
These activity indicators of the primary are consistent with dwarfs of similar spectral types 
in the field and in intermediate age clusters like the Hyades \citep{1999AJ....117..343R, 
2011AJ....141...97W}, and are less active than those of younger open clusters like Pleiades 
\citep{1998ApJ...499L.199S}. 

In the VLT/UVES spectrum of VHS~1256-1257, we did not detect the Li\,{\sc i} resonance doublet 
at 670.8 nm, imposing an upper limit of 30 m\AA~in the pEW of this atomic line (see the 
right panel of Fig.\,\ref{uvesspec_features}). This is much lower than the expected value for 
a full preservation of this element in the atmosphere of late-M dwarfs of EW\,=\,0.5--1\,\AA~\citep{1998ApJ...499L.199S, 
2002A&A...384..937Z}.
The Li element is rapidly destroyed in the interior of stars, on timescales shorter than $\sim$\,150\,Myr, 
and in massive brown dwarfs on timescales of a few Gyr. Brown dwarfs with masses lower than 0.055--0.060 
$M_{\odot}$ do not burn this element in their interiors because their central temperature is not high 
enough to produce this fusion reaction \citep{1998ApJ...497..253U, 2000ApJ...542..464C}. 
The non-detection of Li in VHS~1256-1257 impose a lower limit in the mass of the primary of 0.055--0.060 $M_{\odot}$ 
and also imposes a lower limit to the age of the system, since objects of similar spectral type in the Pleiades 
cluster (age\,$\sim$ 120\,Myr) have fully preserved this element \citep{1998ApJ...499L.199S}. 
Theoretical evolutionary models \citep{1997A&A...327.1039C, 2000ApJ...542..464C} predict
that objects with effective temperature $T_{\rm eff}$\,$\sim$ 2600\,K (which is the corresponding effective temperature 
of an M7.5) have preserved their Li content in timescales of less than 150\,Myr, but have destroyed it for several 
orders of magnitude at ages larger than 200\,Myr.

The gravity-dependent spectral features of 
the primary like Na and K alkaline lines are also consistent with intermediate gravities between those of the 
Pleiades and field dwarfs \citep{2003ApJ...593.1074G, 2013ApJ...772...79A, 2014A&A...562A.127B, 2014MNRAS.439.3890G}
On the other hand, the secondary shows spectral features, which are signposts of low gravity and youth like the 
sharp triangular shape of the $H$ band. Some authors attribute these features to objects with ages younger than 
150\,Myr \citep{2009AJ....137.3345C, 2013MmSAI..84..955F}, which is in contradiction with the lower limit to the age of 
VHS~1256-1257, given by the non-detection of Li in the primary. Other authors suggest a more conservative range 
of ages up to 120--500\,Myr \citep{2014A&A...568A...6Z}, which are in good agreement with the age constrain 
for this system. 

The proper motion, radial velocity, and parallactic distance of the primary allow us to determine its galactic 
kinematic. The galactic velocities UVW of the VHS~1256-1257 indicate that the system probably belongs to the Local 
Association, whose members have estimated ages of 10--300\,Myr \citep{1999A&A...344..376A}. 
In conclusion, based on the absence of lithium in the primary and the likely membership to the Local Association,
we adopt a range of 150--300 M\,yr for the age of the VHS~1256-1257 system.

\begin{figure*}
\centering
 \includegraphics[scale=0.83,keepaspectratio=true]{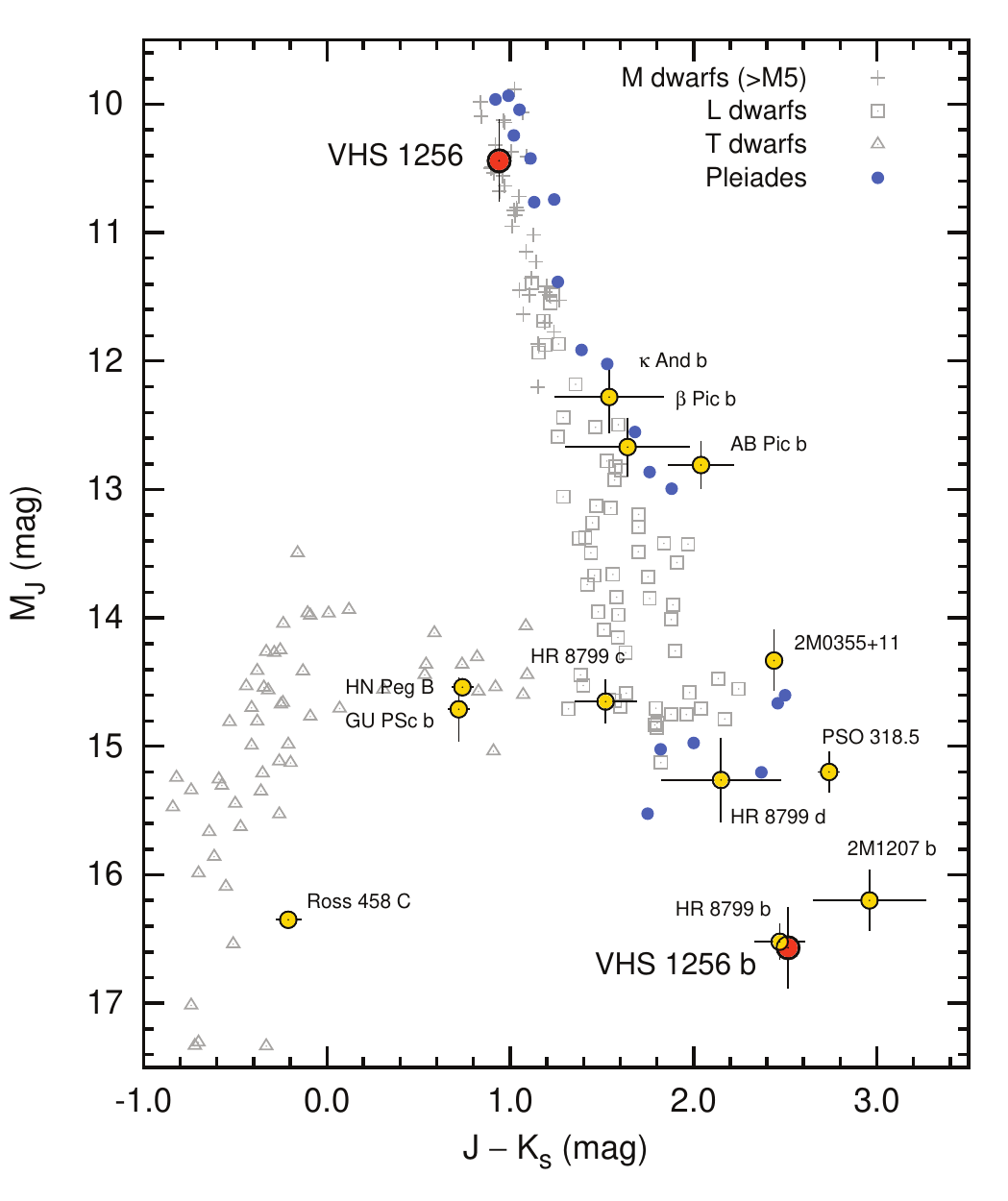}
  \includegraphics[scale=0.83,keepaspectratio=true]{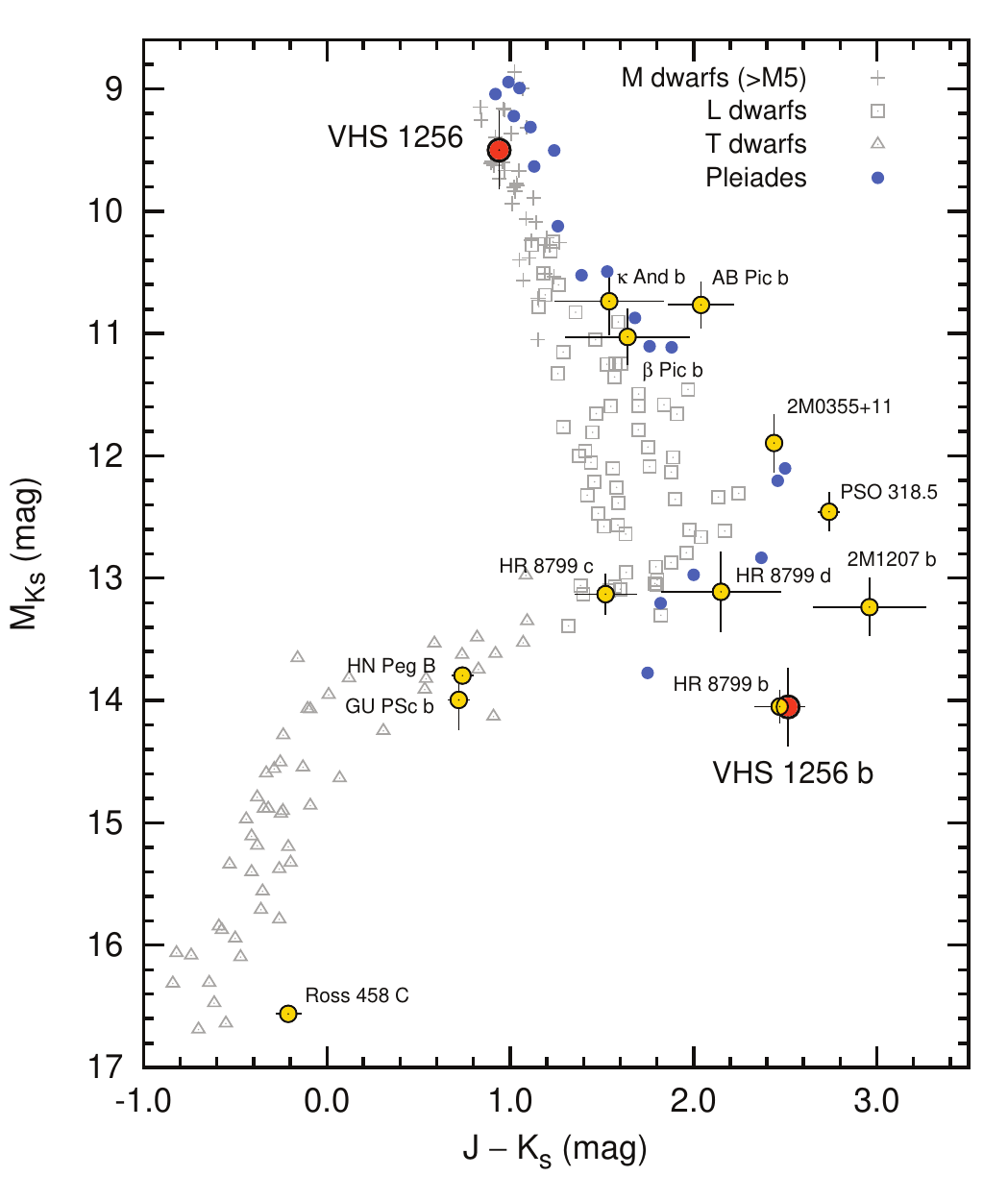}
  \caption{$M_J$ and $M_{Ks}$ vs. $J - K_s$ color-magnitude diagrams comparing the two components of VHS\,1256-1257 with field M, L, 
  and T dwarfs with known parallax measurements from \citet{2012ApJS..201...19D}, known young 
  substellar objects from the compilation of \citet{2013ApJ...774...55B} and the least-massive 
  Pleiades members \citep{2014A&A...568A..77Z}. All photometry was converted to the MKO filter system.}
  \label{cmdiagram}
\end{figure*}

\subsection{$M_{J,\, Ks}$ Versus $J-K_s$ Color-Magnitude Diagrams}
Average near-IR $J-K_s$ color of L7 dwarfs is $1.75\pm0.26$ \citep{2013AJ....145....2F}. With 
$J-K_s$\,=\,2.47\,$\pm$\,0.03~mag, VHS\,1256-1257~b is among the reddest known L dwarfs, next to 2M1207 b and PSO J318.5-22.
These extreme colors are thought to originate from enhanced photospheric dust and
broad-band changes in the spectral energy distribution induced by the low gravity of young objects.
In Figure \ref{cmdiagram} we compare the absolute $J$ and $K_s$ magnitudes versus $J-K_s$ colors of VHS\,1256-1257 with field
mid--late-M, L and T dwarfs with measured parallaxes compiled by \cite{2012ApJS..201...19D} and with several
known substellar objects and giant planet companions.
Based on parallactic distance we derived an absolute magnitude of the companion $M_J=16.45\pm0.30$, which is about
2.2~mag fainter than the late-L field counterparts. In the $K_s$ band it is roughly one magnitude below the LT sequence.
The location of VHS~1256-1257\,b on the $M_J$ and $M_{Ks}$ versus $J-K_s$ color magnitude 
diagrams coincides with the young planetary mass companion 2M~1207\,b and is almost the same 
as the 5--11 $M_{\rm Jup}$ planet HR~8799\,b.

\subsection{Luminosity, Mass, and Effective Temperature}
To derive the bolometric luminosities of the primary and the companion, we used the near-IR 
photometry and the parallactic distance. For the primary, which does not deviate strongly from the field
sequence, we have obtained the bolometric magnitude applying the corresponding BC$_J$ and BC$_K$ bolometric 
corrections determined for field objects, from \citet {2004AJ....127.3516G} and \citet{2002AJ....124.1170D}. 
These BCs combined with absolute magnitudes and $M_{\rm bol}=4.73$ mag for the Sun yield the luminosity 
of $\log(L_{\rm bol}/L_{\odot})=-3.14\pm0.10$ dex. The error accounts for the uncertainties in distance,
photometry, and bolometric correction.
Since the near-IR photometry of the companion differs significantly from the ``normal'' 
field L dwarfs, the bolometric corrections determined for field ultracool dwarfs are not valid 
\citep{2012ARA&A..50...65L, 2012ApJ...752...56F, 2014A&A...568A...6Z}. 
We applied bolometric corrections ($JHK_s$, 2MASS system) derived from the measurements of PSO~J318.5-22 \citep{2013ApJ...777L..20L},
which shows a strong similarity to VHS~1256-1257\,b. Taking the mean value from the three bands,
we obtained $\log(L_{\rm bol}/L_{\odot})=-5.05\pm0.22$ dex.

\begin{figure*}
\centering
  \includegraphics[scale=0.87,keepaspectratio=true]{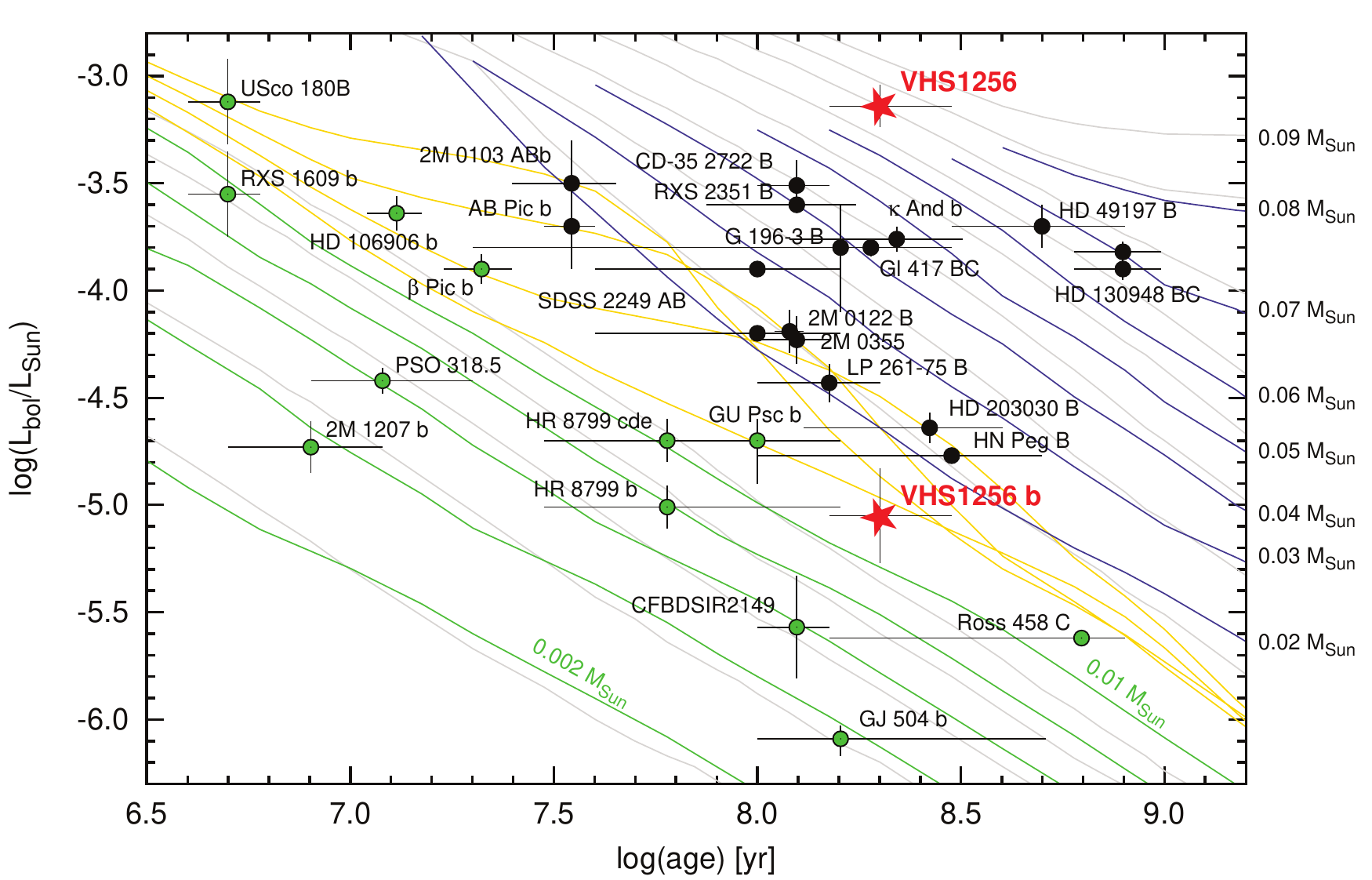}
  \caption{Luminosity and age of the two components of VHS\,1256-1257 compared with evolutionary tracks from the solar 
  abundance, cloudy atmosphere models of \citet{2008ApJ...689.1327S} with $f_{\rm sed}$\,=\,2 and 
  from the BT-Settl models of the Lyon group \citep{2004IAUS..213..119A}. The \citet{2008ApJ...689.1327S} 
  models are plotted in color, with 0.002--0.010~$M_{\odot}$ tracks in green (spaced by 0.002~$M_{\odot}$), 
  0.011--0.014~$M_{\odot}$ tracks in yellow with a 0.001~$M_{\odot}$ step, and 0.02--0.08~$M_{\odot}$ 
  in blue, with 0.01~$M_{\odot}$ increments. The light gray curves correspond to BT-Settl models. For comparison purposes we overplotted 
  the population of known young substellar objects, from the compilations by \citet{2013ApJ...774...55B}, 
  \citet{2013ApJ...777L..20L} and updated with recently discovered objects GU Psc b \citep{2014ApJ...787....5N} 
  and HD~106906 b \citep{2014ApJ...780L...4B}.
  The BT-Settl isomass lines yield slightly higher luminosities at masses $>0.030 M_{\odot}$.
  }
  \label{lum_age}
\end{figure*}

We used the calculated bolometric luminosities to infer the range of possible masses, effective temperatures,
and gravities ($\log\,g$) from the evolutionary models, adopting an age of the system of 150--300 Myr.
We have used the Lyon group models BT-Settl \citep{1998A&A...337..403B, 2003A&A...402..701B, 2000ApJ...542..464C} with 
the \citet{2011SoPh..268..255C} solar abundances and the `hot start' cloudy atmosphere evolutionary tracks 
of \citet{2008ApJ...689.1327S}, with sedimentation parameter $f_{sed}$\,=\,2. 
Both models give consistent values of the derived parameters (masses, $T_{\rm eff}$, $\log\,g$)
within their corresponding uncertainties.
For the primary, we obtained a mass of 73$^{+20}_{-15}$ $M_{\rm Jup}$, close to the boundary between stars and brown dwarfs regime which starts 
at masses below $\sim75$ $M_{\rm Jup}$ for solar metallicities. The effective temperature and $\log\,g$ found
from the models are $2620\pm140$ K and $5.05\pm0.10$~dex, respectively.
This $T_{\rm eff}$ is consistent with typical temperatures of field M$7.5\pm0.5$ dwarfs being in the
$\sim$2500--2600 K range \citep{2000ApJ...535..965L, 2004AJ....127.3516G}.
The derived gravity is slightly lower but similar to the expected gravity for field dwarfs of similar 
spectral types ($\log\,g$\,=\,5.0--5.5). This is consistent with what we have found spectroscopically.

For the companion, we obtained a mass of 11.2$^{+9.7}_{-1.8}$~$M_{\rm Jup}$, indicating that the object 
is near the mass limit at which the onset of deuterium fusion occurs. Given the uncertainty in the 
mass determination, it is currently unclear whether the object is above or below it.
The effective temperature and $\log\,g$ of the companion obtained from the evolutionary models are
880$^{+140}_{-110}$ K and 4.24$^{+0.35}_{-0.10}$ dex. 
A typical effective temperature of field L6--L8 dwarf is between 1600 and 1300 K \citep{2000ApJ...535..965L, 2004AJ....127.3516G, 2004AJ....127.2948V}.
The effective temperature of the companion indicated by the models is about 500 K lower than that and corresponds rather to a field mid-T dwarf. 
A similar discrepancy between temperatures predicted by evolutionary models and those estimated from spectral classification
has been found in other young substellar objects: PSO J318.5-22 \citep{2013ApJ...777L..20L}, 2MASS~0122-2439\,B 
\citep{2013ApJ...774...55B}, HN Peg B \citep{2007ApJ...654..570L}, 2MASS 1207 b \citep{2011ApJ...732..107S, 
2011ApJ...735L..39B}.
Moreover, at this temperature level of $\sim$\,900 K we would already expect to see the methane absorption bands in the near-IR spectra of the companion.
The absence of methane was observed in other similar objects, like for example the HR 8799 planets.
On the contrary, if the actual temperature is higher, like that corresponding to ``normal'' field L6--L8 dwarfs, 
this would lead to unrealistically small radius of the object ($\sim$\,0.5 $R_{\rm Jup}$), given its bolometric luminosity.

We compare in Figure \ref{lum_age} the luminosity of the companion at the adopted range of ages with the 
\citet{2008ApJ...689.1327S} and BT-Settl models \citep{2004IAUS..213..119A} and with presently known young substellar objects, compiled by
\citet{2013ApJ...774...55B} updated with recently reported young T-dwarf companion 
GU PSc b \citep{2014ApJ...787....5N} and a planetary mass companion HD~106906\,b \citep{2014ApJ...780L...4B}. 
Interpretation of evolutionary tracks in this specific region has to be considered 
with caution. At that range of ages and luminosities the onset of deuterium burning causes overlapping 
of isomass tracks and less massive object can be slightly more luminous than a more massive one. This may introduce ambiguities 
in the determination of masses and produce a relatively large uncertainty in the upper limit of the derived mass 
of the companion. This object for its location in the HR diagram and its close distance is an ideal case for
the application of the deuterium test \citep{1999ApJ...521..671B, 2000ApJ...542L.119C}.



\section{Summary and final remarks}
 We have identified an unusually red ($J-K_s$\,=\,2.47 mag) common proper motion L$7\pm1.5$ type companion 
 located at 8\farcs06\,$\pm$\,0\farcs03 ($\sim102$ AU) of an M7.5\,$\pm$\,0.5 dwarf. 
  The near-IR spectrum of the secondary shows a peaked triangular shape of the $H$-band continuum, and 
  other spectral features recognized as hallmarks of low surface gravity and youth. 
 The optical spectrum of the primary shows no Li\,{\sc i} at 670.82 nm at a limit of pEW\,$<$\,30 m\AA. 
 We have determined a parallactic distance of the system of 12.7\,$\pm$\,1.0 pc, which is in agreement with the distance estimated from
 spectral type and photometry of the primary. 
 From the proper motion, distance, and radial velocity of the VHS~1256-1257 
 we obtained the galactic velocities of the primary, which indicate that the system likely belongs to the Local Association.
 The non-detection of lithium and the kinematics of the primary allowed us to constrain the age of the system in the range of 150--300 Myr. 
 From near-IR photometry and bolometric corrections we estimate luminosities of $\log(L_{\rm bol}/L_{\odot})$ 
 of $-3.14\pm0.10$ and $-5.05\pm0.22$~dex for the primary and secondary, respectively.
 By comparison with theoretical evolutionary models we derived a mass of 73$^{+20}_{-15}$ $M_{\rm Jup}$ 
 for the primary, at around the substellar mass limit and 11.2$^{+9.7}_{-1.8}$ $M_{\rm Jup}$ for the secondary, near the deuterium-burning mass limit.
 At the distance of 12.7~pc VHS~1256-1257\,b is among the nearest currently known planetary mass companions detected by direct imaging.
 Moreover, it is one of the very few young, extremely red L dwarfs with age constrained within a narrow range, given by the likely
 belonging to the Local Association and the absence of Li\,{\sc i} in the primary. 
 %

 The $T_{\rm eff}$ of $\sim$\,900 K determined from evolutionary models based on the luminosity does not seem to be 
 consistent with the expected $T_{\rm eff}$ range of field dwarfs of similar spectral type and with the absence of 
 methane, which is expected to appear in atmosphere cooler than 1400 K.
 Following \citet{2011ApJ...733...65B, 2011ApJ...735L..39B} the formation of clouds with substantial vertical thickness and 
 non-equilibrium chemistry in a low-gravity object like VHS~1256-1257\,b could provide an explanation of the apparent high atmospheric
 temperature ($>$1500~K) as compared to cooling track effective temperature predictions (900--1000~K). As atmospheric clouds are 
 composed mostly of Fe and Mg-Si grains, we argue that the infall of planetesimals into this, and other young planetary mass objects, 
 may enrich their atmospheres with key ingredients for substantial cloud formation.
 As time progresses, the amount of infalling material will decrease, the surface gravity of the planet will increase, and therefore 
 the formation of thick clouds will become less important. As a consequence, the majority of field mid L dwarfs will not display 
 the extreme IR colors found in these young objects. An alternative hypothesis to aid the explanation of the very red colors involves
 presence of a warm debris disk or a dust-shell surrounding the object and causing the extinction.
  
 Since it is a relatively nearby and bright object near the deuterium-burning limit, it becomes one of the most 
 promising targets to study the application of the deuterium test. 
 From the masses and separation of the components we estimate the orbital period to be about 3900~yr. Assuming a circular orbit
 with a face-on orientation, the displacement caused by the orbital motion would be from 4 to 13 mas/yr, which will become 
 feasible to measure in the next few years using precise astrometric observations.
\vfill\eject

\acknowledgments
We thank the anonymous referee for his/her valuable comments that improved this manuscript.
B.G. would like to kindly thank Dr. Michael Liu for providing the spectra of PSO J318.5-22.
We thank Roi Alonso and Aitor Bereciartua for performing the IAC80 observations used in this work.
N.L. was funded by the Ram\'on y Cajal fellowship number 08-303-01-02. 
N.L. and V.J.S.B. are financially supported by the projects AYA2010-19136 and AYA2010-20535 
from the Spanish Ministry of Economy and Competitiveness (MINECO), respectively. 
A.P.G. has been supported by Project No. 15345/PI/10 from the Fundaci\'on S\'eneca and MINECO under
the grant AYA2011-29024.
We thank the ESO staff for carrying out the VLT-UVES observations. Allocation of VLT Director’s
Discretionary Time is gratefully acknowledged.
This work is based on observations collected at the European Southern Observatory, Chile, under program numbers
092.C-0874 and 293.C-5014(A).
Based on observation obtained as part of the VISTA Hemisphere Survey, ESO progamme, 179.A-2010 (PI: McMahon). 
The VISTA Data Flow System pipeline processing and science archive are described in \cite{2004SPIE.5493..411I}
and \cite{2009MNRAS.399.1730C}.
This work is based on observations (program GTC65-13B; PI Lodieu) made with the Gran Telescopio Canarias (GTC), 
operated on the island of La Palma in the Spanish Observatorio del Roque de los Muchachos of the Instituto 
de Astrof\'isica de Canarias.
This article is based on observations made with IAC80, NOT, and WHT operated on the islands of Tenerife 
and La Palma by IAC, NOTSA and ING in the Spanish Observatorio del Teide and Roque de los Muchachos.
The data presented here were obtained (in part) with ALFOSC, which is provided by the Instituto de Astrof\'isica 
de Andaluc\'ia (IAA) under a joint agreement with the University of Copenhagen and NOTSA.
This research has benefitted from the Ultracool RIZzo Spectral Library (\url{http://dx.doi.org/10.5281/zenodo.11313}), 
maintained by Jonathan Gagn\'e and Kelle Cruz.

\bibliography{draft}

\end{document}